\documentclass[aps,prd,preprint,groupedaddress,showkeys,amsmath,amssymb]{revtex4-2}

\usepackage{graphicx}
\usepackage{xcolor}
\usepackage{bm}

\newcommand{\nd}{\mathrm{d}}

\newtheorem{lemma}{Lemma}[section]

\begin{document}

% Use the \preprint command to place your local institutional report
% number in the upper righthand corner of the title page in preprint mode.
% Multiple \preprint commands are allowed.
% Use the 'preprintnumbers' class option to override journal defaults
% to display numbers if necessary
%\preprint{}

%Title of paper
\title{Recursive structure of Baikov representations I: Generics and application to symbology}

% repeat the \author .. \affiliation  etc. as needed
% \email, \thanks, \homepage, \altaffiliation all apply to the current
% author. Explanatory text should go in the []'s, actual e-mail
% address or url should go in the {}'s for \email and \homepage.
% Please use the appropriate macro foreach each type of information

\author{Xuhang Jiang}
\email{xhjiang@itp.ac.cn}
\affiliation{CAS Key Laboratory of Theoretical Physics, Institute of Theoretical Physics, Chinese Academy of Sciences, Beijing 100190, China}
\affiliation{School of Physics, Peking University, No.~5 Yiheyuan Road, Beijing 100871, China}
\author{Li Lin Yang}
\email{yanglilin@zju.edu.cn}
\affiliation{Zhejiang Institute of Modern Physics, School of Physics, Zhejiang University, Hangzhou 310027, China}
% \affiliation command applies to all authors since the last
% \affiliation command. The \affiliation command should follow the
% other information
% \affiliation can be followed by \email, \homepage, \thanks as well.
%\author{}
%\email[]{Your e-mail address}
%\homepage[]{Your web page}
%\thanks{}
%\altaffiliation{}
%\affiliation{}

%Collaboration name if desired (requires use of superscriptaddress
%option in \documentclass). \noaffiliation is required (may also be
%used with the \author command).
%\collaboration can be followed by \email, \homepage, \thanks as well.
%\collaboration{}
%\noaffiliation

\date{\today}

\begin{abstract}
    In this paper, we explore the recursive structure of Baikov representations for Feynman integrals. We demonstrate that the various Baikov representations for all sectors of an integral family can be organized in a treelike structure. Using this structure, we show that the symbol letters of one-loop Feynman integrals can be written in terms of minors of a matrix associated with the top sector. Nontrivial relations among these symbol letters can then be easily discovered using results from linear algebra.
\end{abstract}

% insert suggested keywords - APS authors don't need to do this
\keywords{Feynman integrals, Baikov representations, symbol letters}

%\maketitle must follow title, authors, abstract, and keywords
\maketitle

\section{Introduction}
    
    Feynman integrals (FIs) are building blocks of perturbative scattering amplitudes in quantum field theories. To calculate these integrals or to study their properties, we usually write them in a specific representation. In addition to the classical momentum representation and the Feynman (or Schwinger) parametric representation, many new representations have been proposed in the literature. In this work, we focus on the Baikov representation and its generalizations \cite{Baikov:1996iu, Lee:2009dh, Frellesvig:2017aai, Chen:2022lzr}. Because of the complicated integration boundaries of this representation, it is not suitable for direct evaluation. However, the Baikov representations are particularly convenient to study the properties of FIs under cuts, and to study the relations among them. The Baikov representations have been employed to study the integration-by-parts (IBP) relations using methods from algebraic geometry \cite{Larsen:2015ped, Bohm:2018bdy, Kardos:2018uzy, Bendle:2019csk, Chen:2022jux}. They are also useful in the development of the intersection theory for Feynman integrals \cite{Mizera:2017rqa, Mastrolia:2018uzb, Mizera:2019ose, Frellesvig:2019uqt, Frellesvig:2020qot, Weinzierl:2020xyy, Caron-Huot:2021xqj, Caron-Huot:2021iev}, which recast the problem of IBP reduction to the computation of intersection numbers.
    
    An important concept to organize the calculation of Feynman integrals is the so-called uniform transcendentality (UT). UT integrals satisfy $\epsilon$-factorized canonical differential equations \cite{Henn:2013pwa}. In the case that only logarithmic singularities are present, UT integrals can be naturally expressed in terms of multiple polylogarithms \cite{Goncharov:1998kja, Goncharov:2001iea}. Generic methods to construct UT integrals in the Baikov representations have been proposed \cite{Chen:2020uyk, Dlapa:2021qsl, Chen:2022lzr}. The method of symbols \cite{Brown:2009qja, Goncharov:2010jf, Duhr:2011zq} is a very powerful tool to study the analytic and algebraic structures of UT integrals. The symbols contain the information of branch cuts of the integrals as functions of external variables. They also encode various algebraic structures of Feynman integrals, such as shuffle algebras \cite{ree_lie_1958}, stuffle algebras, and Hopf algebras \cite{Goncharov:2005sla, Brown:2011ik, Duhr:2012fh}, as well as cluster algebras \cite{Golden:2013xva, Drummond:2019cxm, Chicherin:2020umh, Mago:2020nuv, Mago:2021luw, Henke:2021ity, He:2021eec, He:2021esx, He:2021non}. The knowledge of the symbol letters can be used to bootstrap the analytic expressions of Feynman integrals (see, for example, \cite{Dixon:2011pw, Dixon:2011nj, Dixon:2016nkn, Drummond:2014ffa, Drummond:2017ssj, Caron-Huot:2020bkp, He:2020vob}).
    
    The Baikov representations take the form of generalized hypergeometric functions
    \begin{equation}
        \int d\bm{x} \, u(\bm{x}) \, \phi(\bm{x}) \, ,
    \end{equation}
    where $\bm{x}$ is the collection of integration variables, $u(\bm{x})$ is a multivalued function determined by the integral family, and $\phi(\bm{x})$ is a rational function representing a specific integral in this family. In both the reduction procedure and the study of symbols, it is important to explore the relations among integrals in all sectors belonging to an integral family. However, the Baikov representations for different sectors can have different numbers of $\bm{x}$ variables, and have different $u(\bm{x})$ functions. This makes it difficult to establish direct relations among them. In this paper, we demonstrate that these different $u(\bm{x})$ functions can be organized in a recursive structure and can be obtained by integrating out variables from the function in the top sector. We further study the symbol letters of one-loop integrals using the canonical differential equations. We find that they can be written in a simple form that explicitly reflects the recursive structure. This can be used to study the analytic structure of one-loop Feynman integrals with arbitrary numbers of external momenta.
    
    The contents are organized as follows. In Sec.~\ref{sec:recursive}, we present the recursive structure of the Baikov representations. In Sec.~\ref{sec:symbol}, we show how one-loop symbol letters are related to this recursive structure, and provide expressions for the symbol letters in a given integral family in terms of the minors of a single matrix. We also show how to derive relations among the letters using the recursive structure, and give some examples. We conclude in Sec.~\ref{sec:conclusion}.

    \section{The recursive structure of Baikov representations}\label{sec:recursive}

    A scalar Feynman integral can be generically written in the momentum representation as
    \begin{equation} \label{eq:FI}
        I(a_1,a_2,\cdots,a_N;d)=\int\frac{\nd^{d}l_1\nd^{d}l_2\cdots\nd^{d}l_L}{(i\pi^{d/2})^L}\frac{1}{x_1^{a_1}x_2^{a_2}\cdots x_N^{a_N}} \, ,
    \end{equation}
    where $L$ is the number of loops and $d=4-2\epsilon$ is the dimension of spacetime; $N=L(L+1)/2+LE$ is the number of independent scalar products involving loop momenta, and $E$ is the number of independent external momenta (the number of external legs is thus $E+1$). The variables $x_{i}$ are propagators if $a_i>0$ and irreducible scalar products (ISPs) if $a_{i}\le 0$.
    
    The momentum representation can be transformed to the standard Baikov representation \cite{Baikov:1996iu, Lee:2010wea}. The derivation was detailed in \cite{Chen:2022lzr}, and we have
    \begin{multline}\label{eq:baikovFI}
        I(a_1,\ldots,a_N;d)=\frac{\pi^{(L-N)/2}\det(A^a_{ij})}{\prod_{i=1}^{L}\Gamma\left(\frac{d-K+i}{2}\right)} \left[ G(p_1,\ldots,p_E) \right]^{-(d-E-1)/2}
        \\
        \times \int\frac{\nd x_1\cdots\nd x_N}{x_1^{a_1}\cdots x_N^{a_N}} \left[ P^{L}_{N}(x_1,\cdots,x_N) \right]^{(d-K-1)/2} \, ,
    \end{multline}
    where $K=L+E$ and the polynomial
    \begin{equation}
        P^{L}_{N}(x_1,\ldots,x_N)=G(q_1,q_2,\ldots,q_K) \, .
    \end{equation}
    A few symbols in the above expressions need to be explained. We use $\{q_1,q_2,\ldots,q_K\}$ to denote $\{l_1,\ldots,l_L,p_1,\ldots,p_E\}$. Any propagator or ISP $x_a$ can be written as a combination of scalar products $q_i \cdot q_j$ ($1 \leq i \leq L$, $i \leq j$) and a term $f_a$ independent of the loop momenta. $A^a_{ij}$ is then the transformation matrix between the variables $\{x_a\}$ and the scalar products $q_i \cdot q_j$
    \begin{equation}
    q_i \cdot q_j = \sum_{a=1}^N A^a_{ij}(x_a-f_a) \, .
    \end{equation}
    $G$ represents the Gram determinant and it can be written for any $n$ momenta $\{q_1,\ldots,q_n\}$ as
    \begin{equation}\label{eq:detQ}
        G(q_1,q_2,\cdots,q_n) \equiv \det M \equiv \det\left(\begin{array}{cccc} q_1^2 & q_1\cdot q_2 & \cdots & q_1\cdot q_n \\ q_2\cdot q_1 & q_2^2 & \cdots & q_2\cdot q_n \\ \vdots & \vdots & \ddots & \vdots \\ q_n\cdot q_1 & q_n\cdot q_2 & \cdots & q_n^2 \end{array}\right) .
    \end{equation}
    The above equation not only defines the Gram determinant, but also defines a symmetric matrix $M$ that will be of crucial importance in the following. It is clear from the definition that if the momenta are redefined by an orthogonal transformation
    \begin{equation}
        (q_1^{\prime},q_{2}^{\prime},\cdots,q_{n}^{\prime}) = O \cdot (q_1,q_2,\cdots,q_n)^{T} \, ,
    \end{equation}
    where $O$ is an orthogonal matrix, the Gram determinant is invariant,
    \begin{equation}\label{eq:graminv}
        G(q_1^{\prime},q_{2}^{\prime},\cdots,q_{n}^{\prime})= G(q_1,q_2,\cdots,q_n) \, .
    \end{equation} 
    We will frequently use this property in later discussions.
    
    In addition to the standard Baikov representation shown above, it is also possible to construct so-called loop-by-loop (LBL) Baikov representations \cite{Frellesvig:2017aai}. The standard way to derive the LBL representations is to perform the change of variables for one loop momentum at a time. On the other hand, it is also possible to derive it from the standard Baikov representation by integrating out some of the Baikov variables \cite{Frellesvig:2017aai, Chen:2022lzr}. In this work, we will explore further this second viewpoint, and show that the various forms of Baikov representations for a given integral family (including all the subsectors) can be cast into a treelike structure rooted in the standard representation. Such a structure allows us to find relations between a sector and its subsectors, which provide important information for integral reductions and differential equations.
    
    The standard and LBL Baikov representations take the generic form 
    \begin{equation}
        \label{eq:gen_rep}
        \int\frac{\nd x_1\cdots\nd x_n}{x_1^{a_1}\cdots x_n^{a_n}} \left[ P_1(\bm{x}) \right]^{\gamma_1}  \cdots \left[ P_m(\bm{x}) \right]^{\gamma_m} \, ,
    \end{equation}
    where we use $\bm{x}$ to denote the sequence of variables $x_1,\ldots,x_n$ with $n \leq N$, and $P_1,\ldots,P_m$ are Baikov polynomials that are raised to noninteger powers $\gamma_1,\ldots,\gamma_m$. In the standard representation, there is of course only one polynomial, i.e., $m = 1$. In the LBL representations, there can be as many as $m = 2L-1$ polynomials for $L$-loop integrals (which we will show). We say that the set of polynomials $\{P_1,\ldots,P_m\}$ \emph{defines} a Baikov representation. When integrating out a variable $x_i$ from the above, one arrives at a different set of Baikov polynomials involving fewer variables. We denote this as a ``lower representation.'' The lower representations usually belong to subsectors, but may also stay at the same sector (in this case, $x_i$ is redundant for this sector). It can also happen that a subsector shares the same representation as its supersector (i.e., no variable is integrated out between them).
    
    We can now state the main result of this section: {the LBL representations for all sectors can be reached by recursively integrating out variables starting from the standard representation of the top-sector}. We will demonstrate this in the following.

\subsection{Behavior of the representation after integrating out a quadratic variable}

    We assume that there exists a variable $x_i$ that only appears quadratically in one of the Baikov polynomials $P_j(\bm{x})$, but not in the other polynomials (in the following we will refer to it as a ``quadratic variable''). We regard $x_i$ as an ISP (i.e., the corresponding power $a_i \leq 0$) either for the current sector or for a subsector. We integrate $x_i$ out to arrive at a lower representation.
    
    The boundary of the integral domain for $x_i$ in Eq.~\eqref{eq:gen_rep} is determined by $P_j(\bm{x})=0$. For simplicity we write $z \equiv x_i$ and
    \begin{equation}
        \label{eq:quad_poly}
        P_j(\bm{x}) = - (A z^2 + B z + C) = -A(z-c_1)(z-c_2) \, ,
    \end{equation}
    where $A$, $B$, and $C$ are polynomials of the remaining variables in $\bm{x}$. We can then integrate $x_i$ out using
    \begin{multline}
        \int_{c_1}^{c_2}z^{n} \left[-A(z-c_1)(z-c_2)\right]^{\gamma}\mathrm{d}z=(-A)^{\gamma} \, c_1^{n} \, (c_2-c_1)^{1+2\gamma} \frac{\Gamma(1+\gamma)^2}{\Gamma(2(1+\gamma))}
        \\
        \times {}_{2}F_{1}\left(1+\gamma,-n,2(1+\gamma); 1-\frac{c_2}{c_1}\right) ,
    \end{multline}
    where $n$ is a non-negative integer.
    By the quadratic transformation and the Pfaff transformation of hypergeometric functions
    \begin{align}
        {}_{2}F_{1}(a,b,2a; z)&=(1-z)^{-\frac{b}{2}} \, {}_{2}F_{1}\left(\frac{b}{2},a-\frac{b}{2},a+\frac{1}{2};\frac{z^2}{4(z-1)}\right) ,
        \\
        {}_{2}F_{1}(a,b,c;z)&=(1-z)^{-a} \, {}_{2}F_{1}\left(a,c-b,c;\frac{z}{z-1}\right) ,
    \end{align}
    we can show that
    \begin{align}
        c_1^{n}\,{}_{2}F_{1}\left(1+\gamma,-n,2(1+\gamma);1-\frac{c_2}{c_1}\right)=\left(\frac{c_1+c_2}{2}\right)^{n}{}_{2}F_{1}\left(-\frac{n}{2},\frac{1-n}{2},\frac{3}{2}+\gamma;\left(\frac{c_1-c_2}{c_1+c_2}\right)^2\right) .
    \end{align}
    So in the end we get the following identity
    \begin{multline}\label{eq:recursionformula}
        \int_{c_1}^{c_2}z^{n} \left[-A(z-c_1)(z-c_2)\right]^{\gamma}\mathrm{d}z=(-A)^{\gamma} \, (c_2-c_1)^{1+2\gamma} \, \frac{\Gamma(1+\gamma)^2}{\Gamma(2(1+\gamma))}
        \\
        \times \left(\frac{c_1+c_2}{2}\right)^{n} \, {}_{2}F_{1}\left(-\frac{n}{2},\frac{1-n}{2},\frac{3}{2}+\gamma;\left(\frac{c_1-c_2}{c_1+c_2}\right)^2\right) .
    \end{multline}
    We will call this identity the {``recursion formula"} hereafter for convenience.
        
    Since $n \ge 0$, the hypergeometric function in the recursion formula is, in fact, a polynomial of $(c_1-c_2)/(c_1+c_2)$. In particular, if $n=0$ or $n=1$, the hypergeometric function just equals $1$. We know that $c_1$ and $c_2$ are the two roots of a quadratic polynomial, which immediately tells us that $c_1+c_2$ and $(c_1-c_2)^2$ are rational functions of the remaining integration variables as well as external variables. Note also that $c_1+c_2$ always appears with positive integer powers in the formula, due to the $(c_1+c_2)^n$ factor in front of the hypergeometric function. Therefore, $c_1+c_2$ simply leads to the polynomial $A$ in the denominator as well as some Baikov variables in the numerator. The latter can be combined with the $x_1^{a_1} \cdots x_n^{a_n}$ denominator, and do not affect the representation. The other rational function $(c_1-c_2)^2$ is just $(B^2-4AC)/A^2$. We will show in the following that $B^2-4AC$ can always be factorized into two polynomials. Therefore, {the effect of integrating out $x_i$ is to replace the polynomial $P_j$ with three new polynomials in the lower representation}.
        
    To show the factorization property of $B^2-4AC$, we start from the standard representation. In this case, there is just one Baikov polynomial $P_1 = \det M$, where the Gram matrix $M$ is defined in \eqref{eq:detQ}. The variable $x_i$ to be integrated out is quadratic in $P_1$. By exploiting properties of the Gram determinant, we can always bring it to the form (up to an overall rational factor)
    \begin{equation}
        \det M \propto \det \tilde{M} \equiv \det
        \begin{pmatrix}
            \times & \times & \times & \cdots & x_i+\times \\
            \times & \times & \times & \cdots & \times \\
            \times & \times & \times & \cdots & \times \\
            \vdots & \vdots & \vdots & \ddots & \vdots\\
            x_i+\times & \times & \times & \cdots & \times
        \end{pmatrix}
        ,
    \end{equation}
    where $\times$ denotes terms independent of $x_i$. The above determinant can be expanded into a quadratic polynomial of the form Eq.~\eqref{eq:quad_poly}, with coefficients written in terms of minors of the matrix $\tilde{M}$. We will write the minors of any matrix $M$ in the form $M_{I,J}$, where $I$ and $J$ are ordered subsequences of the indices $1,2,\ldots,n$. The minor $M_{I,J}$ is defined to be the determinant of the submatrix of $M$ with rows listed in $I$ and columns listed in $J$. For example, $M_{12,\,34}$ is the minor by taking entries in the first or second row and in the third or fourth column. $M_{1,2}$ is simply the $(1,2)$ element of the matrix $M$.
    
    Noting that the discriminant $B^2-4AC$ is invariant under a shift of variable, we have
    \begin{align}
        A &\propto \tilde{M}_{2\ldots n-1,\, 2\ldots n-1} \, , \nonumber
        \\
        B^2-4AC &\propto \left[ \tilde{M}_{2\ldots n,\,1\ldots n-1}^{2} + \tilde{M}_{2\ldots n-1,\,2\ldots n-1} \tilde{M}_{1\ldots n,\,1\ldots n} \right]_{x_i=0} \nonumber
        \\
        &= \tilde{M}_{1\ldots n-1,\,1\ldots n-1} \, \tilde{M}_{2\ldots n,\,2\ldots n} \, . \label{eq:factorize}
    \end{align}
    For the last equal sign we have applied the following identity for an arbitrary symmetric matrix $M$:
    \begin{equation}\label{eq:exchangerelation}
        M_{1X,\,1X} \, M_{Xn,Xn}=M_{1X,\,Xn}^{2} + M_{X,\,X} \, M_{1Xn,\,1Xn} \, ,
    \end{equation}
    where $X$ denotes the sequence $2\ldots n-1$. This identity is usually called Sylvester's identity in the literature \cite{Dlapa:2021qsl, Chen:2022lzr}. It is actually a special case of the more general ``Lewis Carroll identity'' that appears in the discussions of cluster algebras \cite{fomin_introduction_2021}. In our notation it reads 
    \begin{equation}\label{eq:generalexchangerelation}
        M_{X,X^{\prime}} M_{Xab,X^{\prime}cd} = M_{Xa,X^{\prime}c} M_{Xb,X^{\prime}d} - M_{Xa,X^{\prime}d} M_{Xb,X^{\prime}c}\, ,
    \end{equation}
    where $X$ and $X^\prime$ are arbitrary sequences of indices of the same length, and $a,b,c,d$ are four indices absent from $X$ and $X^\prime$. 
    
    From the above, we see that integrating out $x_i$ from $P_1$ produces three polynomials in the lower representation, one being $A$ and the other two from the factorization of $B^2-4AC$. We also note that all the three new polynomials are written as Gram determinants. Hence, it is straightforward to repeat the above procedure for the next quadratic variable, if such a variable exists in the lower representation.

\subsection{The quadratic variables and the recursive structure}
    
    We now analyze how quadratic variables appear in the Baikov polynomials and how they are related to the recursive structure of the Baikov representations for Feynman integrals.
    
    We again start from the standard representation, where there is only one Baikov polynomial. We will show that there is always at least one quadratic variable in this polynomial. It can either be an ISP for the top sector or a propagator in the top sector (and thus an ISP for a subsector). If it is an ISP for the top sector, integrating it out arrives at a new representation for the same sector (quite often the LBL representation). If it is a propagator in the top sector, integrating it out gives a representation for a corresponding subsector.
    
    We now note that the diagrams for all subsectors can be obtained from that of the top sector by ``pinching'' some propagators. These pinched propagators become ISPs for the subsectors (in the standard representation of the top sector). If the Baikov variables corresponding to these ISPs are quadratic, they can be integrated out to arrive at a new representation. Hence, the recursive structure of the Baikov representation is naturally related to the sequence of pinched propagators. The remaining question is then: when does a pinched propagator correspond to a quadratic variable? For that our basic tool is the following Lemma:
    \begin{lemma}
        For an $L$-loop $(E+1)$-point integral sector, if it contains an $(L-1)$-loop subdiagram with the number of external legs less than $E+2$, then there exists at least one quadratic ISP variable in the Baikov polynomial of the standard representation for this sector.
    \end{lemma}
 
    \begin{figure}[t!]
        \centering
        \includegraphics[width=0.3\textwidth]{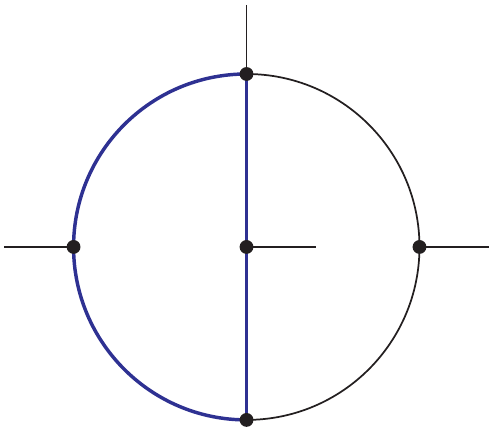}
        \hspace{4em}
        \includegraphics[width=0.3\textwidth]{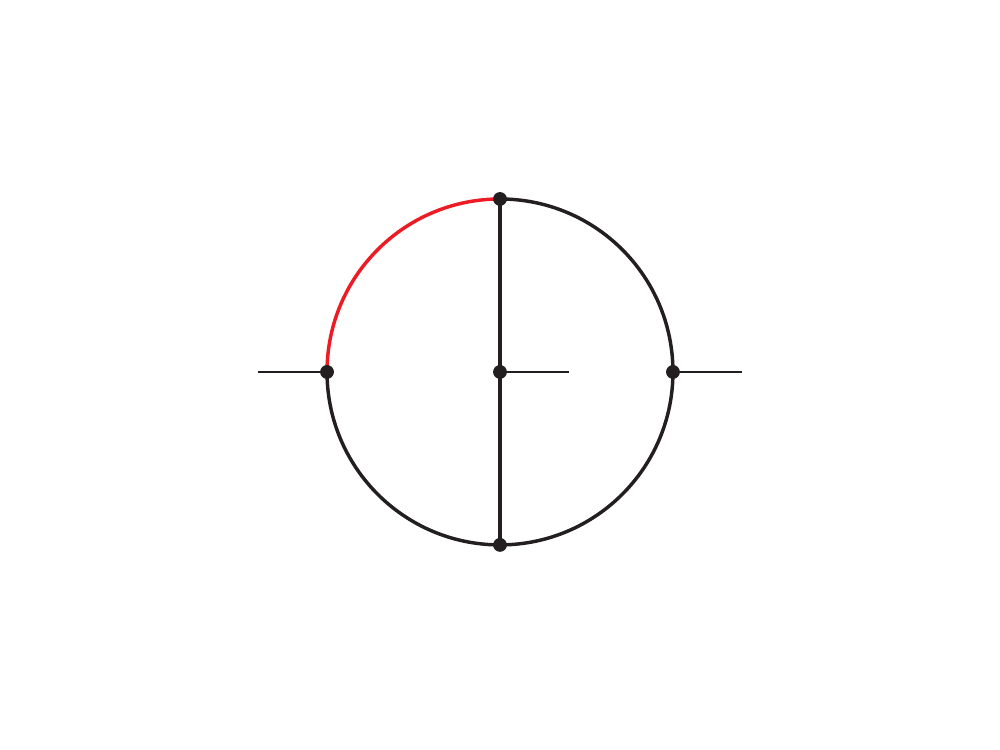}
        \caption{Left: a two-loop, four-point diagram with a quadratic ISP. Right: a two-loop three-point diagram without quadratic ISPs, but the red propagator becomes one after pinching.}\label{fig:simple_examples}
    \end{figure}
      
    It is best to use some simple examples to illustrate the content of the above Lemma. The diagram on the left side of Fig.~\ref{fig:simple_examples} is a two-loop four-point integral sector with a one-loop four-point sub-diagram drawn by blue lines. According to the Lemma, there exists at least one quadratic ISP in the Baikov polynomial. Integrating out this ISP leads to a loop-by-loop representation for this sector. On the other hand, the diagram on the right side of Fig.~\ref{fig:simple_examples} is a two-loop three-point integral sector, with all its one-loop subdiagrams being four-point. It does not satisfy the Lemma, hence no quadratic ISP exists. This means that its loop-by-loop representation is the same as the standard one. However, if one pinches the red propagator, the diagram becomes a two-loop, three-point sector with a one-loop, three-point sub-diagram. The pinched propagator now becomes a quadratic ISP in this subsector, and can be integrated out to arrive at a lower representation.
        
    To show that the Lemma is true, we note that the Baikov polynomial is given by the determinant of the Gram matrix, 
    \begin{equation}\label{eq:gramdet}
        M =
        \begin{pmatrix}
        l_1^2 & \cdots & l_1\cdot l_L & l_1\cdot p_1 & \cdots & l_1\cdot p_E
        \\
        \vdots & \ddots & \vdots & \vdots & \ddots & \vdots
        \\
        l_L \cdot l_1 & \cdots & l_L^2 & l_L\cdot p_1 & \cdots & l_L\cdot p_E
        \\
        p_1\cdot l_1 & \cdots & p_1\cdot l_L & p_1^2 & \cdots & p_1\cdot p_E
        \\
        \vdots & \ddots & \vdots & \vdots & \ddots & \vdots
        \\
        p_E\cdot l_1 & \cdots & p_E\cdot l_L & p_E\cdot p_1 & \cdots & p_E^2
        \end{pmatrix}
        .
    \end{equation}
    The Baikov variables are linear combinations of the scalar products involving the loop momenta $l_i$. In an $(L-1)$-loop subdiagram, one of the loop momenta is treated as an external one. Without loss of generality, we label this loop momentum as $l_L$. In general, this subdiagram can depend on as many as $E+1$ independent external momenta ($l_L$ and $p_1,\ldots,p_E$). However, if the subdiagram satisfies the condition of the Lemma, it can only depend on as many as $E$ independent external momenta (among which $l_L$ must be included). Again, without loss of generality (i.e., after some redefinitions of external momenta), we assume the external momentum dropping out of the subdiagram to be $p_E$, which only appears in the last row/column of the above Gram matrix.
    
    Now, all the propagators of the subdiagram cannot depend on $p_E$, and the remaining propagators of the full diagram must depend on $l_L$. This shows that the $L-1$ scalar products $l_1 \cdot p_E$, ..., $l_{L-1} \cdot p_E$ are ISPs of the full diagram, and they are also quadratic in the Baikov polynomial.
    
    We now turn to the lower representation after integrating out a quadratic variable from the standard representation. As discussed in the last subsection, the Baikov polynomial in the standard representation is replaced by three new polynomials,
    \begin{equation}\label{eq:iterative}
            M_{1Xn,\,1Xn} \rightarrow \left\{ M_{X,\,X} ,\, M_{1X,\,1X} ,\, M_{Xn,\,Xn} \right\} ,
    \end{equation}
    where $n=L+E$ and $X=23\ldots n-1$. All variables contained in $M_{X,\,X}$ must also appear in $M_{1X,\,1X}$ and $M_{Xn,\,Xn}$, and these cannot be quadratic. However, there are scalar products appearing only in the first row/column of $M$ (hence only in $M_{1X,\,1X}$) or only in the last row/column (hence only in $M_{Xn,\,Xn}$). We can then analyze $M_{1X,\,1X}$ and $M_{Xn,\,Xn}$ in the same way as above and locate quadratic ISPs (for the current sector or a subsector) in them. This leads to the recursive structure as expected.
    
    The final question concerns the number of Baikov polynomials in the lower representations. Naively, one might imagine that each recursion step increases the number of polynomials by two. However, the number has an upper bound of $2L-1$. The reason is that during the recursion, certain polynomial factors get cancelled out, restricting the number from increasing indefinitely. As a simple example, suppose that we integrate out $l_1 \cdot p_{E-1}$ from $M_{1X,\,1X}$ in Eq.~\eqref{eq:iterative}. This replaces $M_{1X,\,1X}$ by
    \begin{equation}
            M_{1X,\,1X} \rightarrow \left\{ M_{X^{\prime},\,X^{\prime}} ,\, M_{1X^{\prime},1X^{\prime}} ,\, M_{X,\,X} \right\} ,
    \end{equation}
    where $X^\prime=23\ldots n-2$. The factor of $M_{X,\,X}$ from above exactly cancels that in Eq.~\eqref{eq:iterative} and thus drops out of the lower representation.

    \begin{figure}[t!]
        \centering
        \includegraphics[width=0.6\textwidth]{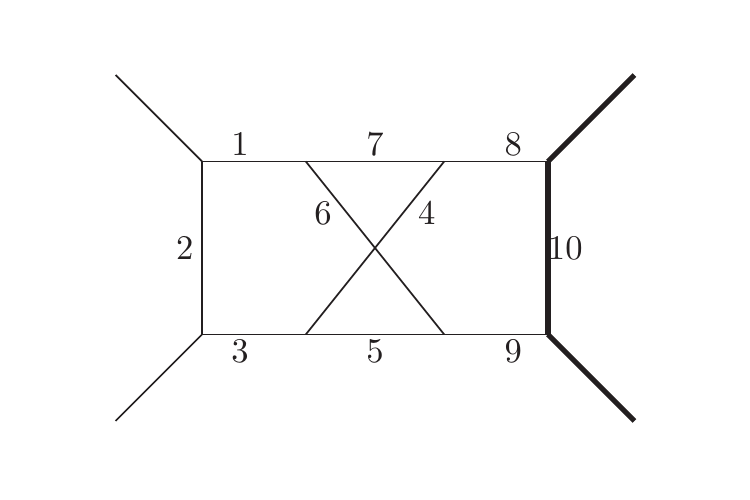}
        \caption{Nonplanar triple-box family. The number beside the internal line is the number of propagators in the definition. All external momenta are assumed incoming and the thick lines are massive with the same mass $m$.}\label{fig:triplebox}
    \end{figure}
    
    We now present a nontrivial example to demonstrate the recursive structure of Baikov representations and their correspondence to the subsectors of the integral family. This three-loop nonplanar triple-box family is defined by the following propagator denominators:
    \begin{multline}
        \{ k_1^2, \, (k_1-p_1)^2,\, (k_1-p_1-p_2)^2, \, k_2^2, \, (k_1-k_2-p_1-p_2)^2, \, (k_1-k_2-k_3)^2, \, \\
        (k_2+k_3)^2, \, k_3^2, \, (k_3-p_1-p_2)^2, \, 
        (k_3-p_1-p_2-p_3)^2-m^2, \, (k_2-p_1)^2, \, \\ (k_2-p_1-p_2)^2, \, (k_2-p_1-p_2-p_3)^2, \, (k_3-p_1)^2, \, (k_1-p_1-p_2-p_3)^2 \} \, .
    \end{multline}
    The kinematics configuration is
    \begin{equation}
        p_1^2=p_2^2=0 \, , \quad p_{3}^2=p_{4}^2=m^2 \, , \quad (p_1+p_2)^2=s \, , \quad (p_2+p_3)^2=t \, .
    \end{equation}
    The diagram for the top sector is depicted in Fig.~\ref{fig:triplebox}, where thick lines represent propagators and external legs with mass $m$, and thin lines are propagators or external legs with zero mass. When picking this integral family, we have in mind the three-loop amplitude for the process $gg\rightarrow t\bar{t}$, although one may assign arbitrary masses to the internal lines without spoiling the recursion. 
    
    \begin{figure}
        \centering
        \includegraphics[width=\textwidth]{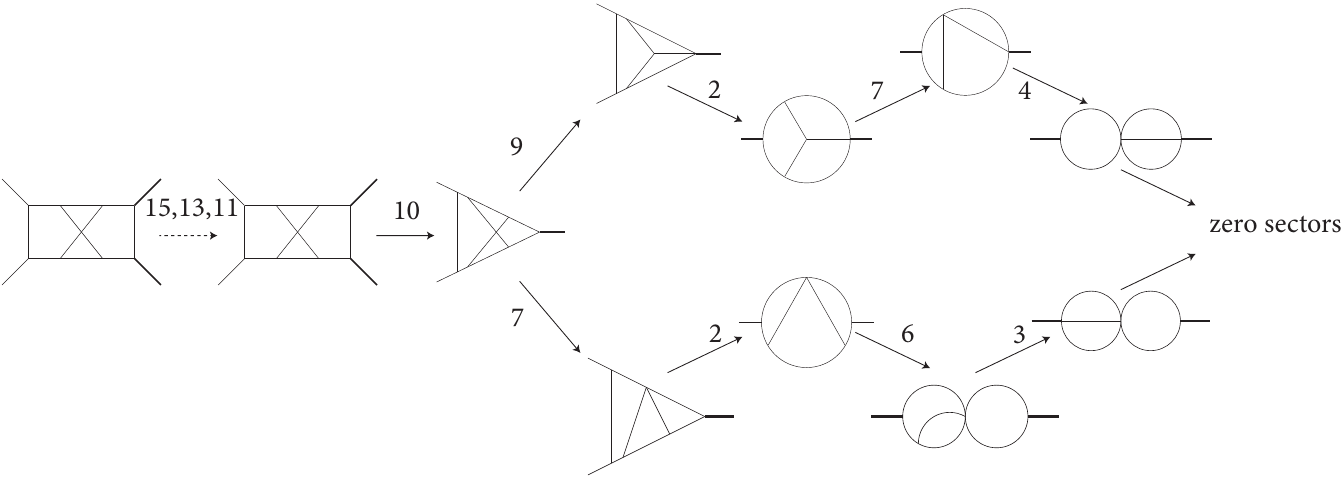}
        %\vspace{-10ex}
        \caption{Two paths in the recursion of Baikov representations for the three-loop nonplanar triple-box family. The variables integrated out are listed next to the arrows. The variables on dashed arrows are ISPs of the current sector, whereas those on solid arrows are ISPs of subsectors.}\label{fig:tripleboxreduce}
    \end{figure}
    
    The whole recursive tree of this family is too large to be shown here. {However, we provide a \texttt{Mathematica} package \texttt{BaikovAll.wl} as Supplemental Material, which can generate all Baikov representations for a given integral family using the recursive structure.} In Fig.~\ref{fig:tripleboxreduce} we show explicitly two paths. The variables integrated out are listed next to the arrows. The variables on dashed arrows are ISPs of the current sector. Integrating them out leads to a lower representation for the same sector. On the other hand, variables on solid arrows are propagators of the current sector. Integrating them out arrives at a representation for a subsector.
    
    While the recursive structure of loop-by-loop Baikov representation might already be familiar to experts in this field, the relations among the various Baikov polynomials in terms of minors of the Gram matrix are, to the best of our knowledge, not present in the literature. In the rest of this paper, we will show that these relations play an important role in the determination of one-loop symbols as well as in the reduction of Feynman integrals.

    \section{Symbol letters from the recursive structure of Baikov representations}\label{sec:symbol}
    
    Usually, the symbol letters can be read off from the $\epsilon$-form differential equations satisfied by UT master integrals. On the other hand, it is highly interesting to obtain the symbol letters without working out the full differential equations. At one-loop level, this has been studied in the projective Feynman parameter representation \cite{Spradlin:2011wp, Arkani-Hamed:2017ahv} and in the Baikov representation \cite{Abreu:2017mtm, Chen:2022fyw, Caron-Huot:2021xqj}. In this section, we revisit this problem from the recursive structure of Baikov representation.

    \subsection{The recursive structure of one-loop Feynman integrals}
    \label{subsec:onelooprecur}
    
    The Baikov polynomial for a one-loop $(E+1)$-point Feynman integral is given by the Gram determinant
    \begin{equation}\label{eq:oneloopgramdet}
        G(l_1,p_1,\ldots,p_{E})=\det\left(\begin{array}{c|ccc} l_1^2 & l_1\cdot p_1 & \ldots & l_1\cdot p_{E} \\   \hline  p_1\cdot l_1 & p_1^2 & \ldots & p_1\cdot p_{E} \\ \vdots &  \vdots & \ddots & \vdots \\ p_{E}\cdot l_1 & p_{E}\cdot p_1 & \ldots & p_{E}^2 \end{array}\right) \, .
    \end{equation}
    We denote the Baikov variables (propagators) by $\{x_1,\ldots,x_{E+1}\}$. They are chosen as
    \begin{multline}
        x_{E+1}=l_1^2-m_1^2 \, , \; x_{E}=(l_1-p_1)^2-m_2^2 \, , \; \ldots \, , \; x_{1}=(l_1-p_1-\cdots-p_{E})^2-m_{E+1}^2 \, .
    \end{multline}
    Note that the choice is completely generic, as one can always relabel the external momentum and internal masses. The choice here reflects a particular order of the recursion.
    We can invert the above relations to express the scalar products $l_1^2$ and $\{l_1 \cdot p_j\}$ in terms of $\{x_i\}$. With the above choice, $x_1$ is only involved in $l_1 \cdot p_E$, $x_2$ is involved in $l_1 \cdot p_E$ and $l_1 \cdot p_{E-1}$, etc. It is then easy to see that $G(l_1,p_1,...,p_E)$ is a quadratic polynomial of the Baikov variables. Hence it can be written in the form
    \begin{equation}\label{eq:generaloneloopP}
        P(\bm{x}) \equiv G(l_1,p_1,\ldots,p_{E}) \equiv \bm{x}\cdot Q \cdot \bm{x}^{T} \, ,
    \end{equation}
    where $Q$ is a symmetric $n \times n$ matrix with $n \equiv E+2$, and the vector $\bm{x}=(x_1,...,x_{E+1},1)$. The elements of $Q$ are functions of $p_{i}\cdot p_{j}$. In the following we will again employ the notation of minors discussed above Eq.~\eqref{eq:factorize}, i.e., $Q_{I,\,J}$ is the minor with the rows in $I$ and the columns in $J$.
    
    As discussed in the last section, we generate new polynomials in the lower representations when integrating out variables one by one. In the following, we will associate these polynomials to the minors of the $Q$ matrix. The latter naturally appears in the expressions of symbol letters. 
    
    Let us integrate out $x_{1}$ first. After the integration, the new Baikov polynomial (at one loop there is always only one Baikov polynomial in each representation) is now a function of $\bm{x}^{(1)} = (x_2,...,x_{E+1},1)$. We can write it (up to an irrelevant constant factor) as
    \begin{equation}\label{eq:P1}
        P_{1}\left(\bm{x}^{(1)}\right)\equiv \bm{x}^{(1)}\cdot Q^{(1)} \cdot \bm{x}^{(1)}{}^{T} \, ,
    \end{equation} 
    with the matrix $Q^{(1)}$ written in terms of minors of $Q$,
    \begin{equation}
        Q^{(1)} = -
        \left(\begin{array}{cccc}
            Q_{12,12} & Q_{12,13} & \cdots & Q_{12,1n} \\
            Q_{13,12} & Q_{13,13} & \cdots & Q_{13,1n}  \\
            \vdots & \vdots & \ddots & \vdots \\
            Q_{1n,12} & Q_{1n,13} & \cdots & Q_{1n,1n}
        \end{array}\right) \, .
    \end{equation}    
    The above result can be easily derived by tracing the quadratic and linear terms of each variable in the original polynomial \eqref{eq:generaloneloopP} and using the results of integration from the last section.
    
    Following the same spirit, we can then integrate out $x_2$, and the Baikov polynomial becomes
    \begin{equation}
        P_{2}\left(\bm{x}^{(2)}\right)\equiv \bm{x}^{(2)}\cdot Q^{(2)} \cdot \bm{x}^{(2)}{}^{T} \, ,
    \end{equation}
    where $\bm{x}^{(2)}=(x_3,x_4,...,1)$ and 
    \begin{equation}
        Q^{(2)} = -
        \left(\begin{array}{cccc}
            Q_{123,123} & Q_{123,124} & \cdots & Q_{123,12n} \\
            Q_{123,124} & Q_{124,124} & \cdots & Q_{124,12n}  \\
            \vdots & \vdots & \ddots & \vdots \\
            Q_{12n,123} & Q_{12n,124} & \cdots & Q_{12n,12n}
        \end{array}\right) \, .
    \end{equation}
    Note that $Q^{(2)}$ is derived from $Q^{(1)}$, but is rewritten in terms of minors of $Q$ using the exchange relations \eqref{eq:exchangerelation}.     
    
    The recursion can further proceed and, after integrating out the first $k$ variables, we arrive at
    \begin{equation}
        P_{k}\left(\bm{x}^{(k)}\right)\equiv \bm{x}^{(k)}\cdot Q^{(k)} \cdot \bm{x}^{(k)}{}^{T} \, ,
    \end{equation}
    where $\bm{x}^{(k)}=(x_{k+1},x_{k+2},...,1)$ and 
    \begin{equation}
        \label{eq:Qk}
        Q^{(k)} = \mathrm{sn}(k)
        \left(\begin{array}{cccc}
            Q_{X(k+1),X(k+1)} & Q_{X(k+1),X(k+2)} & \cdots & Q_{X(k+1),Xn} \\
            Q_{X(k+2),X(k+1)} & Q_{X(k+2),X(k+2)} & \cdots & Q_{X(k+2),Xn}  \\
            \vdots & \vdots & \ddots & \vdots \\
            Q_{Xn,X(k+1)} & Q_{Xn,X(k+2)} & \cdots & Q_{Xn,Xn}
        \end{array}\right) \, ,
    \end{equation}
    where $X = 12\cdots k$, and $\mathrm{sn}(k)$ equals $+1$ if $k$ is a multiple of $3$, and $-1$ otherwise. This sign is introduced for later convenience.
    
    From the above, we see that the Baikov polynomials of all sectors are related to minors of the matrix $Q$. We now employ the method of \cite{Chen:2020uyk, Chen:2022fyw} to construct the $d\log$ integrands for each sector. Setting $N = E+1-k$, the integrands read    
    \begin{align}
        \text{$N$ even} &: g_{k}\left(\bm{x}^{(k)}\right) = \sqrt{\mathrm{sn}(k) \, Q_{Xn,Xn}} \, \left[ \mathrm{sn}(k\!-\!1) \, Q_{X,X} \right]^\epsilon \, \left[ P_k\left(\bm{x}^{(k)}\right) \right]^{-1/2-\epsilon}\frac{1}{x_{k+1} \, x_{k+2} \cdots x_{E+1}} \, , \nonumber
        \\
        \text{$N$ odd} &: g_{k}\left(\bm{x}^{(k)}\right) = \left[ \mathrm{sn}(k-1) \, Q_{X,X} \right]^\epsilon  \left[ P_k\left(\bm{x}^{(k)}\right) \right]^{-\epsilon} \, \frac{1}{x_{k+1} \, x_{k+2} \cdots x_{E+1}} \, , \label{eq:UTbasis}
    \end{align}
    where we have suppressed irrelevant $\bm{x}$-independent factors. These are our starting point to derive the symbol letters.
    
    \subsection{Differential equations and symbol letters}

    To get the symbol letters associated with the $(E+1)$-point UT integral $g_0(\bm{x})$ (which defines the top sector), we need to study its derivatives with respect to external kinematic variables. To this end, we simply take $Q_{i,j}$ as independent variables, and we only need to study $Q_{1,1}$, $Q_{1,2}$, $Q_{1,n}$, and $Q_{n,n}$. The symbol letters related to the other $Q_{i,j}$'s can be obtained by permuting the indices.
    
    We first consider the derivative with respect to $Q_{1,1}$. This gives
    \begin{equation}
        \frac{\partial}{\partial Q_{1,1}} \left[ P(\bm{x}) \right]^{\gamma} \, \frac{1}{x_1x_2 \cdots x_{E+1}} = \left[ P(\bm{x}) \right]^{\gamma-1} \, \frac{x_1^2}{x_1x_2 \cdots x_{E+1}} = \left[ P(\bm{x}) \right]^{\gamma-1} \, \frac{x_1}{x_2x_3 \cdots x_{E+1}} \, ,
    \end{equation}
    where factors independent of $\bm{x}$ have been suppressed, and $\gamma$ can be $(-1/2-\epsilon)$ or $-\epsilon$ depending on whether $E$ is even or odd. Invoking the recursion formula \eqref{eq:recursionformula} to integrate $x_1$ out, the integrand in the lower representation becomes    
    \begin{equation}
        \left[ Q_{1,1} \right]^{-\gamma-1} \left[ P_{1}\left(\bm{x}^{(1)}\right) \right]^{\gamma-1/2} \, \frac{Q_{1,2}x_2+\cdots+Q_{1,E+1}x_{E+1}+Q_{1,n}}{x_2x_3 \cdots x_{E+1}} \, ,
    \end{equation}
    where $P_{1}(\bm{x}^{(1)})$ has been defined in \eqref{eq:P1}. The above integrand is a linear combination of $Q_{1,j}$ with $j > 1$. We conclude that the symbol letters in the derivative with respect to $Q_{1,1}$ are also a linear combination of those obtained from the derivatives with respect to $Q_{1,j}$. Hence, we are left with only $Q_{1,2}$, $Q_{1,n}$, and $Q_{n,n}$ to investigate.
    
    Let us start from the case when $E+1$ is even. Taking the derivative with respect to $Q_{1,2}$, we have
    \begin{align}
        \frac{\partial}{\partial Q_{1,2}} g_0(\bm{x})&=(-1/2-\epsilon)\sqrt{Q_{n,n}} \left[ P(\bm{x}) \right]^{-3/2-\epsilon} \, \frac{1}{x_3x_4 \cdots x_{E+1}} \nonumber \\
        &\rightarrow (-Q_{12,12})^{-1+\epsilon} \left[ P_{2}\left(\bm{x}^{(2)}\right) \right]^{-1/2-\epsilon} \, \frac{\sqrt{Q_{n,n}} \, Q_{12,12}}{x_3x_4 \cdots x_{E+1}} \, .
    \end{align}
    In the second row, we have integrated out $x_1$ and $x_2$ to arrive at the lower representation with $E-1$ propagators. Here and in the following, we use arrows to mean that some variables have been integrated out on the right-hand side. Hereafter, we will only keep track of factors relevant to the symbol letters. Comparing with the first equation of \eqref{eq:UTbasis}, we can easily see that the above integrand is proportional to the $(E-1)$-point UT integrand $g_{2}\left(\bm{x}^{(2)}\right)$,
    \begin{align}\label{eq:partialQ12}
        \frac{\partial}{\partial Q_{1,2}} g_0(\bm{x}) &\to \frac{\sqrt{Q_{n,n}}}{\sqrt{-Q_{12n,12n}}} \, g_{2}\left(\bm{x}^{(2)}\right) \nonumber \\
        &\to \frac{\partial}{\partial Q_{1,2}} \log\frac{Q_{1n,2n}+\sqrt{-Q_{12n,12n}Q_{n,n}}}{Q_{1n,2n}-\sqrt{-Q_{12n,12n}Q_{n,n}}} \, g_{2}\left(\bm{x}^{(2)}\right) \, .
    \end{align}    
    The symbol letter can be read off from the argument of the logarithm.
    
    Now consider the derivative with respect to $Q_{1,n}$,
    \begin{align}\label{eq:partialQ1np1raw}
        \frac{\partial}{\partial Q_{1,n}} g_0(\bm{x}) &= (-1/2-\epsilon) \sqrt{Q_{n,n}} \left[ P(\bm{x}) \right]^{-3/2-\epsilon} \, \frac{1}{x_2x_3 \cdots x_{E+1}} \nonumber \\
        &\rightarrow \left( Q_{1,1} \right)^{-1/2+\epsilon} \left[ P_{1}\left(\bm{x}^{(1)}\right) \right]^{-\epsilon} \, \frac{\sqrt{Q_{n,n}} \, Q_{1,1}}{x_2x_3 \cdots x_{E+1} \, P_{1}\left(\bm{x}^{(1)}\right)} \, .
    \end{align}
    Comparing with the second equation of \eqref{eq:UTbasis}, we see that there is an extra factor of $P_1$ in the denominator. This leads to additional subsector integrals in the relation,
    \begin{align}\label{eq:partialQ1np1}
        \frac{\partial}{\partial Q_{1,n}} g_{0}(\bm{x}) &\to -\frac{\sqrt{Q_{1,1}Q_{n,n}}}{Q_{1n,1n}} \, g_{1}\left(\bm{x}^{(1)}\right) + \text{subsector integrands} \nonumber \\
        &\to \frac{\partial}{\partial Q_{1,n}} \log\frac{Q_{1,n}+\sqrt{Q_{1,1}Q_{n,n}}}{Q_{1,n}-\sqrt{Q_{1,1}Q_{n,n}}} \, g_{1}\left(\bm{x}^{(1)}\right) + \text{subsector integrands}  \, .
    \end{align}
    The first term can be obtained through maximal cut, and the argument of the logarithm gives us another symbol letter. We show in Appendix~\ref{app:explicit} how to obtain the subsector terms by subtraction and that they do not give us new information.

    Finally, we consider the derivative with respect to $Q_{n,n}$,
    \begin{align}
        \frac{\partial}{\partial Q_{n,n}} g_0(\bm{x}) &=\left[ \left( -\frac{1}{2}-\epsilon \right) \sqrt{Q_{n,n}} \left[ P(\bm{x}) \right]^{-3/2-\epsilon} + \frac{1}{2\sqrt{Q_{n,n}}} \left[ P(\bm{x}) \right]^{-1/2-\epsilon} \right] \frac{1}{x_1x_2 \cdots x_{E+1}}  \, .
    \end{align}
    This gives the dependence on $g_0$ itself,
    \begin{align}\label{eq:partialQnp1np1}
        \frac{\partial}{\partial Q_{n,n}} g_0(\bm{x}) &\propto \frac{1}{Q_{n,n}} g_0(\bm{x}) + \text{subsector integrands} \nonumber \\
        &\propto \frac{\partial}{\partial Q_{n,n}} \log Q_{n,n} \, g_0(\bm{x}) + \text{subsector integrands} \, .
    \end{align}
    We can read off a rational letter from the above. Again, the dependence on subsector integrals does not give new information here.
    
    We now turn to the case when $E+1$ is odd. We again consider the derivatives of $g_0$ with respect to $Q_{1,2}$, $Q_{1,n}$ and $Q_{n,n}$. We get the same symbol letters as in the even case, except a new one of the form
    \begin{equation}
        \log \frac{Q_{1,2}+\sqrt{-Q_{12,12}}}{Q_{1,2}-\sqrt{-Q_{12,12}}} \, .
    \end{equation}
    
    In the above, we analyzed the derivatives of the top-sector integral $g_0(\bm{x})$. Owing to the recursive structure of Baikov representations, the same analysis can be readily applied to the derivatives of the subsector integrals $g_k\left(\bm{x}^{(k)}\right)$. We only need to replace the matrix $Q$ by $Q^{(k)}$ and be careful about the extra factor of $Q_{X,X}$ in Eq.~\eqref{eq:UTbasis}. The resulting symbol letters for all sectors can be summarized in the following:
    \begin{itemize}
        \item Rational letters,
              \begin{align}\label{eq:letterrational}
                \log \frac{Q_{Xn,Xn}}{Q_{X,X}}\, .
              \end{align}
        \item Algebraic letters for even $E+1-k$,
              \begin{gather}
                  \log\frac{Q_{Xin,Xjn}+\sqrt{-Q_{Xn,Xn}Q_{Xijn,Xijn}}}{Q_{Xin,Xjn}-\sqrt{-Q_{Xn,Xn}Q_{Xijn,Xijn}}} \,, \quad \log\frac{Q_{Xi,Xn}+\sqrt{Q_{Xi,Xi}Q_{Xn,Xn}}}{Q_{Xi,Xn}-\sqrt{Q_{Xi,Xi}Q_{Xn,Xn}}} \,, \nonumber\\
                  k < i \ne j < n \, . \label{eq:lettereven}
              \end{gather}
        \item Algebraic letters for odd $E+1-k$,
              \begin{gather}
                \log\frac{Q_{Xi,Xj}+\sqrt{-Q_{X,X}Q_{Xij,Xij}}}{Q_{Xi,Xj}-\sqrt{-Q_{X,X}Q_{Xij,Xij}}} \, , \nonumber\\
                k < i \ne j\le n \, . \label{eq:letterodd}
              \end{gather}
    \end{itemize}
    We remind the reader that $X=12\cdots k$ denotes the sequence of missing propagators in the subsector. For the top sector, $X$ is empty and the associated $Q_{X,X}=1$.
    
    An important feature of the above results is that all symbol letters are actually determined by the principal minors of $Q$, because all nonprincipal minors are related to principal minors by
    \begin{equation}
        Q_{Xa,Xb}^{2}=Q_{Xa,Xa}Q_{Xb,Xb}-Q_{X,X}Q_{Xab,Xab} \, .
    \end{equation}
    This exchange relation has also been related to the positivity of minors. If both $Q_{Xa,Xa}Q_{Xb,Xb}$ and $-Q_{X,X}Q_{Xab,Xab}$ are positive, then $Q_{Xa,Xb}$ must be a real quantity. % {\alert [ is this the reason for the $\mathrm{sn}(k)$ factor in the definition of $Q$? ]}

    The symbol letters obtained above can be related to those written in terms of Gram determinants in \cite{Chen:2022fyw}. We give the relations in Appendix~\ref{app:QandGram}. Finally, we note that in the above analysis we have implicitly assumed that $Q_{X,X}\neq 0$ and $Q_{Xn,Xn}\neq 0$. In practice, the situations where $Q_{X,X}=0$ or $Q_{Xn,Xn}=0$ may appear for certain kinematic configurations. The corresponding symbol letters can be obtained from the above generic results by linear combinations followed by a limiting procedure. These have been discussed extensively in \cite{Chen:2022fyw} and we do not repeat them here.

    \subsection{Relations among symbol letters}\label{app:interdependency}
    
    The set of symbol letters obtained above (as well as those obtained using other methods) might be redundant. It is often desirable to find possible relations among the letters to obtain an independent subset. This helps to construct the symbols and to bootstrap the analytic expressions of the integrals. Finding these relations is usually a highly nontrivial task when algebraic letters are present. One can use the program package \texttt{SymBuild} \cite{Mitev:2018kie} for that purpose, but it becomes extremely slow when the letters involve many square roots. An advantage of the $Q$-minor representation is that relations among various letters can be discovered using results in linear algebra. We discuss a set of nontrivial relations in this subsection.
    
    We start from the first letter in Eqs.~\eqref{eq:lettereven} and the letter in \eqref{eq:letterodd}. They can be written in the form
    \begin{equation}
        W_{Yij}=\frac{Q_{Yi,Yj}+\sqrt{-Q_{Y,Y}Q_{Yij,Yij}}}{Q_{Yi,Yj}-\sqrt{-Q_{Y,Y}Q_{Yij,Yij}}} \, ,
    \end{equation}
    where the sequence $Y$ is either $X$ or $Xn$. The minors of $Q$ in the above expression can be reexpressed as minors of the adjugate matrix of $Q$. We denote this adjugate matrix as
    \begin{equation}
        \mathcal{Q} = \left( \det Q \right) Q^{-1} \, ,
    \end{equation}
    and we have 
    \begin{equation}
        W_{Yij}=\frac{\mathcal{Q}_{Y^{\prime}i,Y^{\prime}j}+\sqrt{-\mathcal{Q}_{Y^{\prime},Y^{\prime}}\mathcal{Q}_{Y^{\prime}ij,Y^{\prime}ij}}}{\mathcal{Q}_{Y^{\prime}i,Y^{\prime}j}-\sqrt{-\mathcal{Q}_{Y^{\prime},Y^{\prime}}\mathcal{Q}_{Y^{\prime}ij,Y^{\prime}ij}}} \, ,
    \end{equation}
    where $Y^\prime$ is the sequence complementary to the sequence $Yij$, i.e.,
    \begin{equation}
        Y^{\prime} = \{12 \cdots n\} \setminus \{Yij\} \, .
    \end{equation}
    The minor $\mathcal{Q}_{Y^{\prime},Y^{\prime}}$ can be viewed as a dual representation of $Q_{Yij,Yij}$.
    
    Now we can consider relations among the following three letters
    \begin{equation}
        \begin{aligned}
            W_{Yij}&=\frac{\mathcal{Q}_{Zki,Zkj}+\sqrt{-\mathcal{Q}_{Zk,Zk}\mathcal{Q}_{Zkij,Zkij}}}{\mathcal{Q}_{Zki,Zkj}-\sqrt{-\mathcal{Q}_{Zk,Zk}\mathcal{Q}_{Zkij,Zkij}}} \, , \\
            W_{Yjk}&=\frac{\mathcal{Q}_{Zij,Zik}+\sqrt{-\mathcal{Q}_{Zi,Zi}\mathcal{Q}_{Zijk,Zijk}}}{\mathcal{Q}_{Zij,Zik}-\sqrt{-\mathcal{Q}_{Zi,Zi}\mathcal{Q}_{Zijk,Zijk}}} \, , \\
            W_{Yki}&=\frac{\mathcal{Q}_{Zjk,Zji}+\sqrt{-\mathcal{Q}_{Zj,Zj}\mathcal{Q}_{Zjki,Zjki}}}{\mathcal{Q}_{Zjk,Zji}-\sqrt{-\mathcal{Q}_{Zj,Zj}\mathcal{Q}_{Zjki,Zjki}}} \, ,
        \end{aligned}
    \end{equation}
    where
    \begin{equation}
        Z=\{12 \cdots n\} \setminus \{Yijk\} \, .
    \end{equation}
    To avoid sign ambiguities, we assume $Q_{Zi,Zi},Q_{Zj,Zj},Q_{Zk,Zk}>0$ and $Q_{Zijk,Zijk}<0$. 
    We can rewrite the above three letters as
    \begin{equation}
        \begin{aligned}
            W_{Yij}&=\frac{r_{ij}+\sqrt{-\mathcal{Q}_{Zkij,Zkij}}}{r_{ij}-\sqrt{-\mathcal{Q}_{Zkij,Zkij}}} \, , \\
            W_{Yjk}&=\frac{r_{jk}+\sqrt{-\mathcal{Q}_{Zkij,Zkij}}}{r_{jk}-\sqrt{-\mathcal{Q}_{Zkij,Zkij}}} \, , \\
            W_{Yki}&=\frac{r_{ki}+\sqrt{-\mathcal{Q}_{Zkij,Zkij}}}{r_{ki}-\sqrt{-\mathcal{Q}_{Zkij,Zkij}}} \, ,
        \end{aligned}
    \end{equation}
    where
    \begin{equation}
        \begin{aligned}
            r_{ij}=\frac{\mathcal{Q}_{Zki,Zkj}}{\sqrt{\mathcal{Q}_{Zk,Zk}}}\,, \quad r_{jk}=\frac{\mathcal{Q}_{Zij,Zik}}{\sqrt{\mathcal{Q}_{Zi,Zi}}}\,, \quad r_{ki}=\frac{\mathcal{Q}_{Zjk,Zji}}{\sqrt{\mathcal{Q}_{Zj,Zj}}} \, .
        \end{aligned}
    \end{equation}
    The product of the first two letters is given by
    \begin{equation}
            W_{Yij}W_{Yjk}=\frac{A+B\sqrt{-\mathcal{Q}_{Zijk,Zijk}}}{A-B\sqrt{-\mathcal{Q}_{Zijk,Zijk}}} \, ,
    \end{equation}
    where
    \begin{equation}
        \begin{aligned}
            A&=r_{ij}r_{jk}-\mathcal{Q}_{Zkij,Zkij} \, , \\
            B&=r_{ij}+r_{jk} \, .
            \label{eq:WW_AB}
        \end{aligned}
    \end{equation}
    We will show in the following that, in certain cases, the product $W_{Yij}W_{Yjk}$ reduces to the third letter $W_{Yki}$, and hence the latter is not an independent one.
    
    From the general relations of minors \eqref{eq:generalexchangerelation}, we have the following six identities for an arbitrary square matrix $\Delta$:
    \begin{equation}
        \begin{aligned}
            \Delta_{Zj,Zk}\Delta_{Zjik,Zkij}&=\Delta_{Zji,Zki}\Delta_{Zjk,Zkj}-\Delta_{Zji,Zkj}\Delta_{Zjk,Zki} \, , \\
            \Delta_{Zk,Zi}\Delta_{Zkji,Zijk}&=\Delta_{Zkj,Zij}\Delta_{Zki,Zik}-\Delta_{Zkj,Zik}\Delta_{Zki,Zij} \, , \\
            \Delta_{Zi,Zj}\Delta_{Zikj,Zjki}&=\Delta_{Zik,Zjk}\Delta_{Zij,Zji}-\Delta_{Zik,Zji}\Delta_{Zij,Zjk} \, , \\
            \Delta_{Zi,Zi}\Delta_{Zijk,Zijk}&=\Delta_{Zij,Zij}\Delta_{Zik,Zik}-\Delta_{Zij,Zik}^2 \, , \\
            \Delta_{Zj,Zj}\Delta_{Zjki,Zjki}&=\Delta_{Zjk,Zjk}\Delta_{Zji,Zji}-\Delta_{Zjk,Zji}^2 \, , \\
            \Delta_{Zk,Zk}\Delta_{Zkij,Zkij}&=\Delta_{Zki,Zki}\Delta_{Zkj,Zkj}-\Delta_{Zki,Zkj}^2 \, ,
        \end{aligned}
    \end{equation}
    where $Z$ is a sequence of indices not including $i$, $j$, and $k$. Solving the above identities, the following triple product relation can be derived:
    \begin{multline}\label{eq:tripleproduct}
        \Delta_{Zi,Zj}\Delta_{Zj,Zk}\Delta_{Zijk,Zijk}=\Delta_{Zi,Zj}\Delta_{Zij,Zjk}\Delta_{Zjk,Zki}+\Delta_{Zj,Zk}\Delta_{Zki,Zij}\Delta_{Zij,Zjk}\\ +\Delta_{Zj,Zj}\Delta_{Zij,Zki}\Delta_{Zki,Zjk} \, .
    \end{multline}
    Further relations can be obtained by permuting the indices. These relations are simplified when $\Delta_{Z,Z}=0$. In this case, we have the additional relations
    \begin{equation}\label{eq:equalty}
        \begin{aligned}
            \Delta_{Zi,Zj}\Delta_{Zj,Zk}&=\Delta_{Zi,Zk}\Delta_{Zj,Zj}\,,  \\
            \Delta_{Zj,Zk}\Delta_{Zk,Zi}&=\Delta_{Zj,Zi}\Delta_{Zk,Zk}\,,  \\
            \Delta_{Zk,Zi}\Delta_{Zi,Zj}&=\Delta_{Zk,Zj}\Delta_{Zi,Zi} \, ,
        \end{aligned}
    \end{equation}
    from which the following identities can be derived:
    \begin{equation}
        \Delta_{Zi,Zj}^{2}=\Delta_{Zi,Zi}\Delta_{Zj,Zj}\,, \quad \Delta_{Zj,Zk}^{2}=\Delta_{Zj,Zj}\Delta_{Zk,Zk} \,, \quad \Delta_{Zk,Zi}^{2}=\Delta_{Zk,Zk}\Delta_{Zi,Zi} \, .
    \end{equation}
    We now set $\Delta = \mathcal{Q}$ and assume $\mathcal{Q}_{Z,Z}=0$, so the triple product relation \eqref{eq:tripleproduct} becomes
    \begin{equation}
        \mathcal{Q}_{Zijk,Zijk}=\frac{\mathcal{Q}_{Zij,Zjk}\mathcal{Q}_{Zjk,Zki}}{\mathcal{Q}_{Zj,Zk}}+\frac{\mathcal{Q}_{Zki,Zij}\mathcal{Q}_{Zij,Zjk}}{\mathcal{Q}_{Zi,Zj}}+\frac{\mathcal{Q}_{Zjk,Zki}\mathcal{Q}_{Zki,Zij}}{\mathcal{Q}_{Zk,Zi}} \, .
    \end{equation}
    With the sign configuration $\mathcal{Q}_{Zi,Zj}<0$ $\mathcal{Q}_{Zj,Zk}<0$ and $\mathcal{Q}_{Zk,Zi}>0$, we finally have
    \begin{equation}
        \mathcal{Q}_{Zijk,Zijk}=-r_{ki}r_{ij}-r_{jk}r_{ki}+r_{ij}r_{jk} \, .
    \end{equation}
    Plugging the above into Eq.~\eqref{eq:WW_AB}, we have $A = r_{ki} B$, which leads to
    \begin{equation}
        W_{Yij}W_{Yjk} = W_{Yki} \, ,
        \label{eq:relation_letters}
    \end{equation}
    i.e., the three letters are not independent when $\mathcal{Q}_{Z,Z}=0$.
    
    The situation where $\mathcal{Q}_{Z,Z} = Q_{Yijk,Yijk} = 0$ can happen in triangle integrals and lower point integrals. For example, consider that after pinching a sequence $X$ of propagators we arrive at a triangle subsector with external momenta $q_1$, $q_2$, and $q_3$ (they are combinations of the original external momenta $\{p_i\}$). We have
    \begin{equation}
        Q_{Xn,Xn} = \det\left(\begin{array}{ccc} 0 & q_{1}^{2} & q_{3}^{2} \\   q_{1}^{2} & 0 & q_{2}^{2} \\ q_{3}^2 & q_{2}^2 & 0  \end{array}\right) .
    \end{equation}
    Apparently, if there exists one $q_i^2 = 0$, the above determinant vanishes, and the condition $\mathcal{Q}_{Z,Z}=0$ is satisfied. We show a simple example here. Consider the massless hexagon integral family ($E=5$, $n=7$). After pinching the propagators with indices $X=123$, we get a triangle diagram with a massless external leg. From the above, we immediately know that $Q_{1237,1237} = \mathcal{Q}_{456,456} = 0$. We hence obtain a relation among the three letters $W_{Y12}$, $W_{Y23}$, and $W_{Y13}$ according to Eq.~\eqref{eq:relation_letters}, with $Y=7$ here. There are more relations when considering different subsectors. The symbol letters $\log W$ then form a linear system, from which we can solve for the independent ones. We will discuss these more explicitly in the next subsection.

    \subsection{Examples of one-loop symbol letters}\label{subsec:symbolexample}
    
    The generic results \eqref{eq:letterrational} $\sim$ \eqref{eq:letterodd} of one-loop symbol letters can be easily applied to an integral family. Because of the recursive structure, the symbol letters for all sectors can be computed from the minors of the $Q$ matrix of the top sector, which can be straightforwardly programmed. In the following, we give a few examples to demonstrate our method.

    \begin{figure}[t!]
        \centering
        \includegraphics[width=0.6\textwidth]{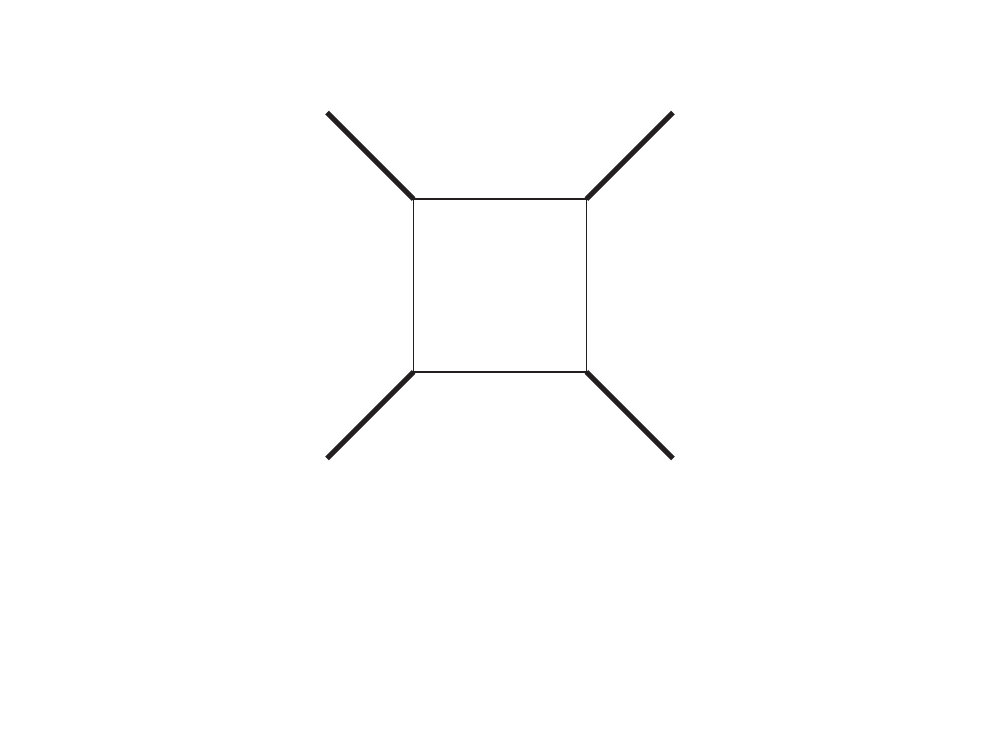}
        \vspace{-15ex}
        \caption{The four-mass-box integral family. The masses of the four external legs are different from each other.}\label{fig:fourmassbox}
    \end{figure}
        
    The first example is the four-mass-box integral family, which has been considered in \cite{He:2022ctv} and we will follow their notations here. The diagram is depicted in Fig.~\ref{fig:fourmassbox}. The propagator denominators are given by
    \begin{equation}
        \{l_{1}^2 , \, (l_{1}-p_{1})^2, \, (l_{1}-p_{1}-p_{2})^2, \, (l_{1}-p_{1}-p_{2}-p_{3})^2\} \, ,
    \end{equation}
    with the kinematic configuration
    \begin{equation}
        p_{i}^2=m_{i}^2 \;\; (i=1,2,3,4) \,, \quad (p_1+p_2)^2=s\,, \quad (p_2+p_3)^2=t \, . 
    \end{equation}
    The entries of the corresponding $5\times 5$ $Q$ matrix are given by
    \begin{eqnarray}\label{eq:fourmassboxQ}
            Q_{1,1}&=-2 m_2^2 t-2 m_3^2 t+m_2^4-2 m_3^2 m_2^2+m_3^4+t^2 \, , \nonumber \\
            Q_{1,2}&=m_3^2 s-m_2^2 s+m_3^2 t-m_4^2 t-m_3^4-2 m_1^2 m_3^2+m_2^2 m_3^2+m_4^2 m_3^2+m_2^2 m_4^2+s t \, , \nonumber \\
            Q_{1,3}&=m_1^2 t+m_2^2 t+m_3^2 t+m_4^2 t-m_1^2 m_2^2+m_1^2 m_3^2+m_2^2 m_4^2-m_3^2 m_4^2-2 s t-t^2 \, , \nonumber \\
            Q_{1,4}&=m_2^2 s-m_3^2 s+m_2^2 t-m_1^2 t-m_2^4+m_1^2 m_2^2+m_3^2 m_2^2-2 m_4^2 m_2^2+m_1^2 m_3^2+s t \, , \nonumber \\
            Q_{1,5}&=m_2^2 s t+m_3^2 s t-2 m_3^2 m_2^2 t+m_4^2 m_2^2 t+m_1^2 m_3^2 t-m_4^2 m_2^4+m_1^2 m_3^2 m_2^2+m_3^2 m_4^2 m_2^2 \nonumber \\
                    &-m_1^2 m_3^4-s t^2 \, , \nonumber \\
            Q_{2,2}&=-2 m_3^2 s-2 m_4^2 s+m_3^4-2 m_4^2 m_3^2+m_4^4+s^2\, , \nonumber \\
            Q_{2,3}&= m_4^2 s-m_1^2 s+m_4^2 t-m_3^2 t-m_4^4+m_1^2 m_4^2-2 m_2^2 m_4^2+m_3^2 m_4^2+m_1^2 m_3^2+s t \, , \nonumber \\
            Q_{2,4}&= m_1^2 s+m_2^2 s+m_3^2 s+m_4^2 s+m_1^2 m_3^2-m_2^2 m_3^2-m_1^2 m_4^2+m_2^2 m_4^2-s^2-2 s t \, , \nonumber \\
            Q_{2,5}&= m_3^2 s t+m_4^2 s t+m_1^2 m_3^2 s-2 m_4^2 m_3^2 s+m_2^2 m_4^2 s-m_1^2 m_3^4+m_1^2 m_4^2 m_3^2+m_2^2 m_4^2 m_3^2 \nonumber \\
                    &-m_2^2 m_4^4-s^2 t \, , \nonumber 
    \end{eqnarray}
    \begin{eqnarray}
            Q_{3,3}&=-2 m_1^2 t-2 m_4^2 t+m_1^4-2 m_4^2 m_1^2+m_4^4+t^2 \, , \nonumber \\
            Q_{3,4}&=m_1^2 s-m_4^2 s+m_1^2 t-m_2^2 t-m_1^4+m_2^2 m_1^2-2 m_3^2 m_1^2+m_4^2 m_1^2+m_2^2 m_4^2+s t \, , \nonumber \\
            Q_{3,5}&=m_1^2 s t+m_4^2 s t+m_3^2 m_1^2 t-2 m_4^2 m_1^2 t+m_2^2 m_4^2 t-m_3^2 m_1^4+m_2^2 m_4^2 m_1^2+m_3^2 m_4^2 m_1^2 \nonumber \\
                    &-m_2^2 m_4^4-s t^2 \, , \nonumber \\
            Q_{4,4}&=-2 m_1^2 s-2 m_2^2 s+m_1^4-2 m_2^2 m_1^2+m_2^4+s^2 \, , \nonumber \\
            Q_{4,5}&=m_1^2 s t+m_2^2 s t-2 m_2^2 m_1^2 s+m_3^2 m_1^2 s+m_2^2 m_4^2 s-m_3^2 m_1^4+m_2^2 m_3^2 m_1^2+m_2^2 m_4^2 m_1^2 \nonumber \\
                    &-m_2^4 m_4^2-s^2 t \, , \nonumber \\
            Q_{5,5}&=-2 m_1^2 m_3^2 s t-2 m_2^2 m_4^2 s t+m_1^4 m_3^4-2 m_1^2 m_2^2 m_4^2 m_3^2+m_2^4 m_4^4+s^2 t^2 \, .
    \end{eqnarray}
    
    We can now apply our method to obtain all possible symbol letters for this family. For the rational letters, we get the same expressions as $W_1,\ldots,W_{10},W_{13},W_{18}$ in \cite{He:2022ctv}. The independent algebraic letters are given by
    \begin{align}
            A_{1}&=\frac{h_1+r_1}{h_1-r_1}\,, \quad A_{2}= \frac{h_2+r_1}{h_2-r_1}\,, \quad A_{3}=\frac{h_3+r_2}{h_3-r_2}\,, \quad A_{4}=\frac{h_4+r_2}{h_4-r_2}\,, \nonumber \\
            A_{5}&=\frac{h_5+r_3}{h_5-r_3}\,, \quad A_{6}= \frac{h_6+r_3}{h_6-r_3}\,, \quad A_{7}=\frac{h_7+r_4}{h_7-r_4}\,, \quad A_{8}=\frac{h_8+r_4}{h_8-r_4}\,, \nonumber \\
            A_{9}&=\frac{h_9+r_5}{h_9-r_5}\,, \quad A_{10}= \frac{h_{10}+r_5}{h_{10}-r_5} \,, \quad A_{11}=\frac{Q_{4,5}+r_1r_2}{Q_{4,5}-r_1r_2}\,, \quad A_{12}= \frac{Q_{3,5}+r_1r_3}{Q_{3,5}-r_1r_3}\,, \nonumber \\
            A_{13}&= \frac{Q_{2,5}+r_1r_4}{Q_{2,5}-r_1r_4}\,, \quad A_{14}= \frac{Q_{1,5}+r_1r_5}{Q_{1,5}-r_1r_5}\,,
    \end{align}
    where the expressions of $r_i$ were given in \cite{He:2022ctv}, and they are actually related to minors of the $Q$ matrix,
    \begin{equation}
        r_1^2=Q_{5,5}\,, \quad r_2^2=Q_{4,4}\,, \quad r_3^2=Q_{3,3}\,, \quad r_4^2=Q_{2,2}\,, \quad r_5^2=Q_{1,1} \, .
    \end{equation}
    The expressions of $h_i$ are given by
    \begin{equation}
        \begin{aligned}
            h_1 &= m_1^2 m_3^2-m_2^2 m_4^2-s t\,, \quad h_2 = -m_1^2 m_3^2-m_2^2 m_4^2+s t\,, \\
            h_3 &= m_1^2-m_2^2-s\,, \quad h_4 = -m_1^2-m_2^2+s\,, \\
            h_5 &= m_1^2-m_4^2-t\,, \quad h_6 = -m_1^2+m_4^2-t\,, \\
            h_7 &= m_3^2-m_4^2-s\,, \quad h_8 = -m_3^2+m_4^2-s\,, \\
            h_9 &= m_2^2-m_3^2-t\,, \quad h_{10} = -m_2^2-m_3^2+t \,,
        \end{aligned}
    \end{equation}
    and they are all related to the minors of the $Q$ matrix. These algebraic letters are equivalent to $W_{21},\ldots,W_{30},W_{46},W_{62},W_{51},W_{59}$ in \cite{He:2022ctv}.

    \begin{figure}[t!]
        \centering
        \includegraphics[width=0.35\textwidth]{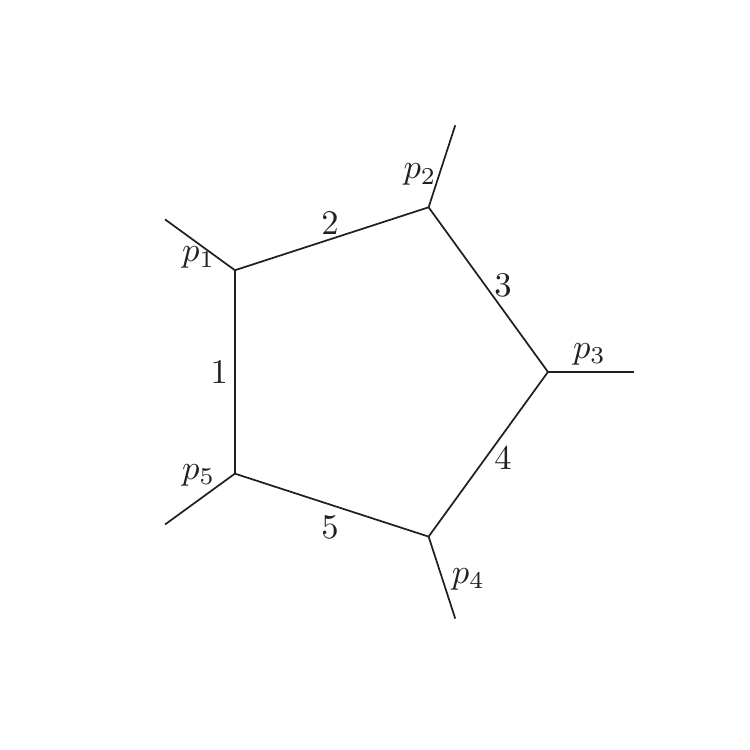}
        \caption{The massless pentagon integral family. All the propagators and external legs are massless. The serial numbers of propagators are labeled aside.}\label{fig:masslesspentagon}
    \end{figure}
    
    The next example is the massless pentagon integral family depicted in Fig.~\ref{fig:masslesspentagon}. The canonical differential equation of this diagram has been well studied in \cite{Kozlov:2015kol,Chicherin:2020oor}. Here we only discuss the symbol letters arising from the dependence of the pentagon integral in the top sector on the box and triangle integrals in the subsectors. Here and in the following, we use the notation $W_X$ to represent the symbol letter corresponding to the subsector after pinching the propagators listed in the sequence $X$.
    There are five different box subsectors, and we can easily get the corresponding symbol letters from our formula \eqref{eq:letterodd}. They can be written as
    \begin{equation}
        \begin{aligned}
            W_{X}=\frac{-\sqrt{\Delta}-R_{X}}{\sqrt{\Delta}-R_{X}}\,, \;\; X=1,2,3,4,5 \, .
        \end{aligned}
    \end{equation}
    where $\Delta \equiv 16G(p_1,p_2,p_3,p_4)$ and
    \begin{equation}
        \begin{aligned}
            R_{1}&=s_{12} s_{15}+s_{45} s_{15}-s_{12} s_{23}+s_{23} s_{34}-s_{34} s_{45} \, , \\
            R_{2}&=s_{12} s_{15}-s_{45} s_{15}+s_{12} s_{23}-s_{23} s_{34}+s_{34} s_{45} \, , \\
            R_{3}&=s_{12} s_{15}-s_{45} s_{15}-s_{12} s_{23}-s_{23} s_{34}+s_{34} s_{45} \, , \\
            R_{4}&=s_{12} s_{15}-s_{45} s_{15}-s_{12} s_{23}+s_{23} s_{34}+s_{34} s_{45} \, , \\
            R_{5}&=s_{12} s_{15}-s_{45} s_{15}-s_{12} s_{23}+s_{23} s_{34}-s_{34} s_{45} \, .
        \end{aligned}
    \end{equation}
    The dependence of the pentagon integral on the ten triangles can be derived from \eqref{eq:letterodd} in a similar way. The corresponding symbol letters are
    \begin{align}
            W_{12}&=\frac{-\sqrt{\Delta}-R_{4}-2s_{23}s_{45}}{\sqrt{\Delta}-R_{4}-2s_{23}s_{45}}\,, \quad W_{13}=\frac{-\sqrt{\Delta}(s_{15}-s_{23})-R_{13}}{\sqrt{\Delta}(s_{15}-s_{23})-R_{13}}\,, \nonumber \\
            W_{14}&=\frac{-\sqrt{\Delta}(s_{15}-s_{34})-R_{14}}{\sqrt{\Delta}(s_{15}-s_{34})-R_{14}}\,, \quad W_{15}=\frac{-\sqrt{\Delta}+R_{3}-2s_{12}s_{34}}{\sqrt{\Delta}+R_{3}-2s_{12}s_{34}}\,, \nonumber \\
            W_{23}&=\frac{-\sqrt{\Delta}+R_{5}-2s_{15}s_{34}}{\sqrt{\Delta}+R_{5}-2s_{15}s_{34}}\,, \quad W_{24}=\frac{-\sqrt{\Delta}(s_{12}-s_{34})-R_{24}}{\sqrt{\Delta}(s_{12}-s_{34})-R_{24}}\,, \nonumber \\
            W_{25}&=\frac{-\sqrt{\Delta}(s_{12}-s_{45})-R_{25}}{\sqrt{\Delta}(s_{12}-s_{45})-R_{25}}\,, \quad W_{34}=\frac{-\sqrt{\Delta}-R_{1}-2s_{12}s_{45}}{\sqrt{\Delta}-R_{1}-2s_{12}s_{45}}\,, \nonumber \\
            W_{35}&=\frac{-\sqrt{\Delta}(s_{23}-s_{45})-R_{35}}{\sqrt{\Delta}(s_{23}-s_{45})-R_{35}}\,, \quad W_{45}=\frac{-\sqrt{\Delta}-R_{2}-2s_{15}s_{23}}{\sqrt{\Delta}-R_{2}-2s_{15}s_{23}} \, ,
    \end{align}
    where
    \begin{align}
            R_{13}&=s_{12} s_{15}^2-s_{45} s_{15}^2-2 s_{12} s_{23} s_{15}+s_{23} s_{34} s_{15}+s_{23} s_{45} s_{15}+s_{34}s_{45} s_{15}+s_{12} s_{23}^2-s_{23}^2 s_{34} \nonumber \\
            &+s_{23} s_{34} s_{45}\, , \nonumber \\
            R_{14}&=s_{12} s_{15}^2-s_{45} s_{15}^2-s_{12} s_{23} s_{15}-s_{12} s_{34} s_{15}-s_{23} s_{34} s_{15}+2 s_{34}s_{45} s_{15}+s_{23} s_{34}^2-s_{12} s_{23} s_{34} \nonumber \\
            &-s_{34}^2 s_{45} \, , \nonumber \\
            R_{24}&=s_{15} s_{12}^2-s_{23} s_{12}^2-s_{15} s_{34} s_{12}+2 s_{23} s_{34} s_{12}-s_{15} s_{45} s_{12}-s_{34}s_{45} s_{12}-s_{23} s_{34}^2+s_{34}^2 s_{45} \nonumber \\
            &-s_{15} s_{34} s_{45} \, , \nonumber \\
            R_{25}&=s_{15} s_{12}^2-s_{23} s_{12}^2+s_{23} s_{34} s_{12}-2 s_{15} s_{45} s_{12}+s_{23} s_{45} s_{12}+s_{34}s_{45} s_{12}+s_{15} s_{45}^2-s_{34} s_{45}^2 \nonumber \\ 
            &+s_{23} s_{34} s_{45} \, , \nonumber \\
            R_{35}&=-s_{12} s_{23}^2+s_{34} s_{23}^2+s_{12} s_{15} s_{23}+s_{12} s_{45} s_{23}+s_{15} s_{45} s_{23}-2 s_{34}s_{45} s_{23}-s_{15} s_{45}^2+s_{34} s_{45}^2 \nonumber \\
            &+s_{12} s_{15} s_{45} \, .
    \end{align}
    
    One can observe that the above letters involve only one square root: $\sqrt{\Delta}$. This indicates that there may exist extra relations among them. This is indeed the case. When we pinch two propagators in the massless pentagon diagram, we always get a triangle diagram with at least one massless external leg. This satisfies the condition $\mathcal{Q}_{Z,Z}=0$ discussed in the previous subsection Sec.~\ref{app:interdependency}. As a result, we find that all the $W_{ij}$ letters in the above can be generated from the five letters $W_1,\ldots,W_5$, which are the truly independent ones. For example, $W_{1}$ and $W_{2}$ will produce $W_{12}$, etc.

    \begin{figure}[t!]
        \centering
        \includegraphics[width=0.4\textwidth]{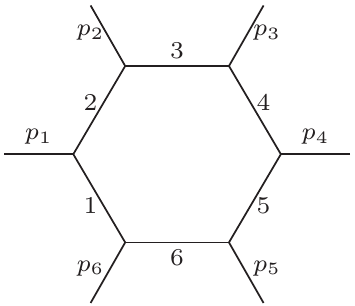}
        \caption{The massless hexagon integral family. All the external legs and propagators are massless. The serial numbers of propagators are labeled aside.}\label{fig:masslesshexagon}
    \end{figure}
    
    Finally, we consider the massless hexagon integral family depicted in Fig.~\ref{fig:masslesshexagon}. The hexagon symbol letters in $d=6$ dimensions have been discussed in \cite{DelDuca:2011ne, Spradlin:2011wp}. For $d=4-2\epsilon$ under dimensional regularization, there are more letters than the integer-dimension case. They have been considered in \cite{Henn:2022ydo}. We will only discuss letters that come from the dependence of the hexagon integral in the top sector on the pentagon and box integrals in the subsectors. In this case, there are 6 pentagons and 15 boxes, and naive counting indicates that there should be 21 symbol letters. However, we find that only nine of them are independent. Among them are the six letters associated with the dependence on the pentagon integrals,
    \begin{equation}
            W_{X}=\frac{-\sqrt{R_{2} \, S_{X}}-T_{X}}{\sqrt{R_{2} \, S_{X}}-T_{X}} \,, \quad X=1,\ldots,6 \, .
    \end{equation}
    The lengthy expressions of $S_{X}$ and $T_{X}$ are given in the Supplemental Material \footnote{See Supplemental Material at http://link.aps.org/supplemental/10.1103/PhysRevD.108.076004 for the package BaikovAll.wl and the notebook UsageofBaikovAll.nb, as well as a .m file for explicit expressions of hexagon's symbol letters discussed in the main text.}. Note that the six $S_X$'s are all different, and there exist no further relations among these six letters. On the other hand, the 15 letters associated with the box integrals involve only one square root: $\sqrt{R_{2}}$. This is, of course, no coincidence and is again related to the condition $\mathcal{Q}_{Z,Z}=0$ (with many different choices of $Z$) discussed in the previous example. This generates a lot of relations, from which we solve that only 3 out of the 15 letters are independent. They can be chosen as, e.g., $W_{12}$, $W_{23}$, and $W_{34}$.
    
    We emphasize that the additional relations generated under the condition $\mathcal{Q}_{Z,Z}=0$ are associated with massless external legs. When some of the external legs become massive, fewer relations can exist and there will be more independent symbol letters.

    \section{Conclusion and Outlook}\label{sec:conclusion}

    In this paper, we have surveyed the recursive structure existing in the Baikov representations for the various sectors of an integral family. Starting from the standard Baikov representation of an integral family, we can derive the other Baikov representations for all sectors in this family by integrating out Baikov variables recursively. This leads to a treelike structure that allows us to analyze the relations among integrals in different sectors. We employ this structure to study the appearance of subsector integrals in the derivatives of a chosen Feynman integral. We find that we can reconstruct all the one-loop symbol letters from the derivatives without performing contour integrals like in \cite{Abreu:2017mtm, Chen:2022fyw}. The letters can be written in terms of the minors of a single matrix, which directly reflects the recursive structure of the Baikov representations. This unified representation of the letters allows us to study the relations among them in a systematic way, utilizing the algebraic identities among the minors. These identities can be used to determine the independent symbol letters in a problem, which helps bootstrap the analytic solutions for the integrals.
    
    The most interesting finding in this paper is that the information of all sectors in an integral family (at arbitrary loops) is contained in a single matrix. It will be interesting to investigate whether this structure at higher-loop orders can also help to reconstruct the symbol letters in a way similar to the one-loop case. It is not that straightforward due to the fact that the representations usually involve more than one Gram determinant, and hence cannot always be written in the form of Eq.~\eqref{eq:generaloneloopP}. Nevertheless, for integral families admitting $d\log$-form integrands \cite{Chen:2020uyk, Chen:2022lzr}, it is still possible to read off the information about the symbol letters.{ We have explored several nontrivial examples, and will present the higher-loop extension in a forthcoming article.}
    
    In addition to the symbology, integral reduction can also benefit from the knowledge about the interconnection among integrands in different sectors within a family. In the reduction procedure, one needs to solve a large linear system involving integrals in all sectors. This may require an extreme amount of computational resources for complicated problems. Alternatively, one may employ the top-down approach, where one first performs the reduction under maximal cut and then moves to the subsectors with the top-sector subtracted integrands. In a forthcoming article, we will demonstrate how the recursive structure can help the top-down reduction of Feynman integrals.

\begin{acknowledgments}
    We would like to thank Song He, Jiaqi Chen and Xiaofeng Xu for useful discussions and collaborations on related subjects.
    This work was supported in part by the National Natural Science Foundation of China under Grants No. 11975030 and No. 12147103 and the Fundamental Research Funds for the Central Universities.
\end{acknowledgments}

\appendix
    
    \section{Relations between $Q$ minors and Gram determinants}\label{app:QandGram}
    
    In this work, we have written the one-loop symbol letters in terms of minors of the $Q$ matrix. On the other hand, in \cite{Chen:2022fyw} they are written in terms of Gram determinants of external momenta. In this appendix, we discuss the relations between the two representations.
    
    First of all, from Eqs.~\eqref{eq:oneloopgramdet} and \eqref{eq:generaloneloopP}, one can easily see that
    \begin{equation}
        \label{eq:Q11}
        Q_{1,1} = -\frac{1}{4} G(p_1,p_2,\ldots,p_{E-1}) \, .
    \end{equation}
    The other $Q_{i,i}$ for $i < n = E+2$ can be obtained by permuting and relabeling the external momenta. As for $Q_{n,n}$, it simply equals $P(\bm{x})$ in Eq.~\eqref{eq:generaloneloopP} with all $x_i = 0$. Using Eq.~\eqref{eq:graminv}, we can reorganize the entries in the determinant to obtain
    \begin{multline}\label{eq:Qnp1}
        Q_{n,n}= \\ \det\left(\begin{array}{cccc} m_{E+1}^{2} & (m_{E}^2-m_{E+1}^2-p_{1}^{2})/2 & \cdots & (m_{1}^{2}-m_{E+1}^{2}-p_{12\cdots E}^{2})/2 \\   (m_{E}^2-m_{E+1}^2-p_{1}^{2})/2 & p_1^2 & \cdots & p_1\cdot p_{12\cdots E} \\ \vdots &  \vdots & \ddots & \vdots \\ (m_{1}^{2}-m_{E+1}^{2}-p_{12\cdots E}^{2})/2 & p_{12\cdots E}\cdot p_1 & \cdots & p_{12\cdots E}^2 \end{array}\right) \, ,
    \end{multline}
    where $p_{X} \equiv \sum_{i \in X} p_i$ for the sequence $X$ of indices. When all propagators are massless, i.e., $m_{i}=0$, Eq.~\eqref{eq:Qnp1} reduces to
    \begin{align}\label{eq:Qspecialform}
        Q_{n,n}&= \det\left(\begin{array}{cccc} 0 & -p_{1}^{2}/2 & \cdots & -p_{12\cdots E}^{2}/2 \\   -p_{1}^{2}/2 & p_1^2 & \cdots & p_1\cdot p_{12\cdots E} \\ \vdots &  \vdots & \ddots & \vdots \\ -p_{12\cdots E}^{2}/2 & p_{12\cdots E}\cdot p_1 & \cdots & p_{12\cdots E}^2 \end{array}\right) \nonumber \\
        &= \left(-\frac{1}{2}\right)^{E+1} \det\left(\begin{array}{ccccc} 0 & p_{1}^{2} & p_{12}^{2} & \cdots & p_{1\cdots E}^{2} \\   p_{1}^{2} & 0 & p_{2}^{2} & \cdots & p_{2\cdots E}^{2}\\ p_{12}^2 & p_{2}^2 & 0 & \cdots & p_{3\cdots E}^2 \\ \vdots & \vdots & \vdots & \ddots & \vdots \\ p_{1\cdots E}^{2} & p_{2\cdots E}^2 & p_{3\cdots E}^2 & \cdots & 0 \end{array}\right) \nonumber
        \\
        &=  G(l_1,p_1,\ldots,p_E)\Big|_{\text{maximal cut}} \, .
    \end{align}
    Note that the second line in the above equation is particularly useful in the momentum-twistor representation.
    
    We now turn to the principal minors $Q_{X,X}$ and $Q_{Xn,Xn}$ with more rows/columns. Recall that $X$ is the sequence of indices corresponding to the integrated-out variables. The resulting minors appear as entries of the matrix $Q^{(k)}$ in Eq.~\eqref{eq:Qk}. To relate them to Gram determinants, we can follow the recursion presented in Sec.~\ref{sec:recursive}. According to the discussions around Eq.~\eqref{eq:factorize}, after integrating out $x_{1}$ from the top sector, we arrive at the factors $G(p_1,\ldots,p_{E-1})$ and
    \begin{multline}
        G(p_1,\ldots,p_{E}) \, G(l_1,p_1,\ldots,p_{E-1})=  G(p_1,\ldots,p_{E})\times \\ \det\left(\begin{array}{c|ccc} l_1^2 & l_1\cdot p_1 & \ldots & l_1\cdot p_{E-1} \\   \hline  p_1\cdot l_1 & p_1^2 & \ldots & p_1\cdot p_{E-1} \\ \vdots &  \vdots & \ddots & \vdots \\ p_{E-1}\cdot l_1 & p_{E-1}\cdot p_1 & \ldots & p_{E-1}^2 \end{array}\right) .
        \label{eq:GGGG}
    \end{multline}
    Integrating out $x_2$ from the above cancels the $G(p_1,\ldots,p_{E-1})$ factor and modifies the second factor in Eq.~\eqref{eq:GGGG} to $G(l_1,p_1,\ldots,p_{E-2})$. Similar behavior occurs when integrating out further variables. Therefore, taking into account the correct prefactors, we have the relations
        \begin{equation}
            \begin{aligned}
                Q_{X,X}&=\left(-\frac{1}{4}\right)^{\mathrm{dim}X}\frac{G(\bm{p}_{X})}{G(p_1,\ldots,p_{E})} \, , 
                \\
                Q_{Xn,Xn}&=\left(-\frac{1}{4}\right)^{\mathrm{dim}X}\frac{G(\bm{q}_{X})|_{\text{maximal cut}}}{G(p_1,\ldots,p_{E})} \, ,
            \end{aligned}
        \end{equation}
        where $\bm{p}_X$ denotes the set of independent external momenta after pinching the propagators in $X$, and $\bm{q}_X$ contains in addition the loop momentum. Apparently, the above results agree with Eq.~\eqref{eq:Q11} for $X=1$ and with Eq.~\eqref{eq:Qspecialform} for $X=\emptyset$\footnote{$G(p_{1},\ldots,p_{E})$ in the denominator originates from absorbing the kinematic factor of the top sector into $P_{x}$ defined in \eqref{eq:generaloneloopP}. All quantities are rescaled by this factor so we can ignore it when it is not important.}. We have checked that our results of one-loop symbol letters agree with those in \cite{Chen:2022fyw} using the above relations.

    \begin{figure}[t!]
        \centering
        \includegraphics[width=0.4\textwidth]{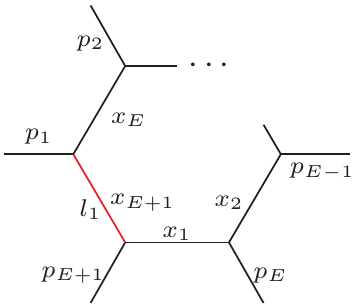}
        \caption{The momenta configuration of a one-loop $(E+1)$-point diagram. It helps to figure out the momenta configurations after pinching $x_2$ and $x_{E+1}$.}\label{fig:sketch}
    \end{figure}
    
    As a simple example, take $X=2,E+1$, and after pinching (see Fig.~\ref{fig:sketch} for reference), the momenta configurations become
    \begin{equation}
        \begin{aligned}
            \bm{p}_{X}&=\{p_2,\ldots,p_{E-2},p_{E-1}+p_{E}\} \, , \\
            \bm{q}_{X}&=\{l_1+p_1,p_2,\ldots, p_{E-2},p_{E-1}+p_{E}\} \,.
        \end{aligned}
    \end{equation}
    Note that since $x_{E+1}=l_1^2$ is pinched, we need to use $l_1+p_1$ as loop momentum.

    \section{Subsector integrands in Eq.~\eqref{eq:partialQ1np1}}\label{app:explicit}
    
    In this appendix, we explicitly show that the subsector integrands in Eq.~\eqref{eq:partialQ1np1} do not generate new symbol letters. The subsector terms are given by
    \begin{equation}\label{eq:subtopology}
        \text{subsector integrands} = \left( Q_{1,1} \right)^{-1/2+\epsilon} \left[ P_{1}\left(\bm{x}^{(1)}\right) \right]^{-\epsilon}
        \frac{\sqrt{Q_{n,n}} \, Q_{1,1}}{Q_{1n,1n} \, P_{1}(\bm{x}^{(1)})} \, \frac{P_{1}(\bm{x}^{(1)})+Q_{1n,1n}}{x_2x_3 \cdots x_{E+1}} \, ,
    \end{equation}
    where the numerator can be written as
    \begin{equation}\label{eq:numerator}
        P_{1}(\bm{x}^{(1)})+Q_{1n,1n} = -\sum_{i,j=2}^{n-1} Q_{1i,1j} \, x_i x_j - 2\sum_{i=2}^{n-1} Q_{1i,1n} \, x_i \, .
    \end{equation}
    The terms linear in $x_i$ will generate dependence on some $g_{2}(\bm{x}^{(2)})$, which is already covered in Eq.~\eqref{eq:partialQ12}\footnote{Here we have used the fact that $Q_{1i,1n}$ (the coefficient of $x_i$ in Eq.~\eqref{eq:numerator}) is a function of $Q_{1,i}$. Therefore this term is already contained in $\partial_{Q_{1,i}}g_{0}(\bm{x})$.}. The quadratic terms will generate dependence on some $g_{3}(\bm{x}^{(3)})$, which seems to be new. In the following we show that these new contributions actually vanish.
    
	Without loss of generality, we consider the dependence of $\partial_{Q_{1,n}} g_{0}(\bm{x})$ on the $g_{3}(\bm{x}^{(3)})$ with $x_1$, $x_2$, and $x_3$ integrated out,
	\begin{equation}
	    \label{eq:g3}
        g_{3}(\bm{x}^{(3)})= \left[ -Q_{123,123}\right]^{\epsilon} \left[ P_3(\bm{x}^{(3)}) \right]^{-\epsilon}  \frac{1}{x_4 \cdots x_{E+1}} \, .
    \end{equation}
    The relevant terms in \eqref{eq:numerator} are
    \begin{equation}
        \label{eq:numerator123}
        -Q_{12,12}x_2^2-Q_{13,13}x_3^2-2Q_{12,13}x_2x_3 \, .
    \end{equation}
    Let us begin with the first term $-Q_{12,12}x_2^2$. After cancelling $x_2$ in the denominator, we integrate $x_2$ out using the recursion formula. We then get
    \begin{equation}
        \left[ -Q_{12,12} \right]^{\epsilon} \left[ P_2(\bm{x}^{(2)} \right]^{-\epsilon-1/2}  
        \frac{\sqrt{Q_{n,n}}}{Q_{1n,1n}} \, \frac{Q_{12,13} \, x_3 + \cdots}{x_3x_4 \cdots x_{E+1}} \, .
    \end{equation}
    The ellipsis represents terms independent of $x_3$. They are irrelevant to our discussion here. Taking the first term $Q_{12,13} \, x_3$ and integrating out $x_3$, we arrive at
    \begin{equation}
        \left[ -Q_{123,123} \right]^{\epsilon-1/2} \left[ P_3(\bm{x}^{(3)} \right]^{-\epsilon} \frac{\sqrt{Q_{n,n}}Q_{12,13}}{Q_{1n,1n}} \, .
    \end{equation}
    Comparing with Eq.~\eqref{eq:g3}, we see that the above expression is just $c_1 g_{3}(\bm{x}^{(3)})$, with the coefficient
    \begin{equation}
        c_1=\frac{Q_{12,13}\sqrt{Q_{n,n}}}{Q_{1n,1n}\sqrt{-Q_{123,123}}} \, .
    \end{equation}
    Because of the symmetry between $x_2$ and $x_3$ in the above analysis, the second term $-Q_{13,13}x_3^2$ in Eq.~\eqref{eq:numerator123} generates the same coefficient $c_2=c_1$. Finally, for the last term $-2Q_{12,13}x_2x_3$, we can directly integrate out $x_2$ and $x_3$ to get
    \begin{equation}
        \left[ -Q_{123,123} \right]^{\epsilon-1/2} \left[ P_3(\bm{x}^{(3)} \right]^{-\epsilon} \frac{-2\sqrt{Q_{n,n}}Q_{12,13}}{Q_{1n,1n}} \equiv c_3 g_{3}(\bm{x}) \, ,
    \end{equation}
    with the coefficient
    \begin{equation}
        c_3=-2\frac{Q_{12,13}\sqrt{Q_{n,n}}}{Q_{1n,1n}\sqrt{-Q_{123,123}}} \, .
    \end{equation}
    Therefore we have
    \begin{equation}
        c_1+c_2+c_3=0 \, .
    \end{equation}
    Hence, we conclude that the subsector integrands in Eq.~\eqref{eq:partialQ1np1} do not lead to new symbol letters.
    
    The above conclusion can be made more generic, that the derivatives of $g_{0}(\bm{x})$, in general, do not depend on $g_{3}(\bm{x})$. Exceptions can happen when some $g_{2}(\bm{x})$ is reducible, and hence is not a true master integral. In these cases the dependence on this $g_{2}(\bm{x})$ is carried over to lower-point integrals. This has also been discussed in \cite{Chen:2022fyw}.

\bibliographystyle{apsrev4-2}
\bibliography{ref}

\end{document}